\documentclass[12pt,preprint,showpacs,preprintnumbers,nofootinbib,%
tightenlines,superscriptaddress,amsmath,amssymb]{revtex4}

\newcommand{\PRE}[1]{{#1}} 

\usepackage{amssymb,color}
\usepackage{graphicx}
\usepackage{dcolumn}
\usepackage{bm}
\usepackage{subfigure}
\usepackage{hyperref}

\usepackage{color}
\usepackage[normalem]{ulem}

\def\beqn{\begin{eqnarray}} 
\def\eeqn{\end{eqnarray}} 
\def\be{\begin{equation}}
\def\ee{\end{equation}}

\def\beq{\begin{eqnarray}}
\def\eeq{\end{eqnarray}}
\def\bea{\begin{eqnarray}}
\def\eea{\end{eqnarray}}

\newcommand{\gev}{\text{GeV}}
\newcommand{\tev}{\text{TeV}}
\newcommand{\zb}{\text{zb}}
\newcommand{\pb}{\text{pb}}

\newcommand{\cm}{\text{cm}}

\newcommand{\km}{\text{km}}

\newcommand{\sr}{\text{sr}}
\newcommand{\s}{\text{s}}
\newcommand{\yr}{\text{yr}}

\newcommand{\eg}{{\em e.g.}}

\newcommand{\eqnref}[1]{Eq.~(\ref{#1})}

\newcommand{\secref}[1]{Sec.~\ref{sec:#1}}

\newcommand{\figref}[1]{Fig.~\ref{fig:#1}}
\newcommand{\figsref}[2]{Figs.~\ref{fig:#1} and \ref{fig:#2}}
\newcommand{\Figref}[1]{Figure~\ref{fig:#1}}

\newcommand{\mgaugino}{M_{1/2}}

\newcommand{\drbar}{\ensuremath{\overline{\mbox{\sc dr}}}}

\newcommand{\hthreem}{{\sc H3m}}
\newcommand{\FeynHiggs}{{\sc FeynHiggs}}

\newcommand{\sigmaSI}{\sigma_p^{\text{SI}}}
\newcommand{\sigmaSD}{\sigma_p^{\text{SD}}}

\begin{document}

\preprint{UCI-TR-2013-05}
\preprint{HU-EP-13/14}
\preprint{SFB/CPP-13-21}

\PRE{\vspace*{0.7in}}

\title{Dark Matter Detection in Focus Point Supersymmetry
\PRE{\vspace*{0.3in}} }

\author{Patrick Draper}
\email{pidraper@ucsc.edu} 
\affiliation{Department of Physics, University of California, 1156
  High Street, Santa Cruz, CA 95064, USA}
\affiliation{Santa Cruz Institute for Particle Physics, Santa Cruz, CA
  95064, USA}

\author{Jonathan L. Feng}
\affiliation{Department of Physics and Astronomy, University of
  California, Irvine, CA 92697, USA}

\author{Philipp Kant}
\email{philipp.kant@physik.hu-berlin.de} 
\affiliation{Humboldt-Universit\"at zu Berlin, 12489 Berlin, Germany}

\author{Stefano Profumo}
\email{profumo@ucsc.edu} 
\affiliation{Department of Physics, University of California, 1156
  High Street, Santa Cruz, CA 95064, USA}
\affiliation{Santa Cruz Institute for Particle Physics, Santa Cruz, CA
  95064, USA}

\author{David Sanford
\PRE{\vspace*{.3in}}}
\email{dsanford@caltech.edu} 
\affiliation{California Institute of Technology, Pasadena, CA 91125,
  USA
\PRE{\vspace*{.5in}}}

\begin{abstract}
\PRE{\vspace*{.2in}} We determine the prospects for direct and
indirect detection of thermal relic neutralinos in supersymmetric
theories with multi-TeV squarks and sleptons.  We consider the
concrete example of the focus point region of minimal supergravity,
but our results are generically valid for all models with decoupled
scalars and mixed Bino-Higgsino or Higgsino-like dark matter.  We
determine the parameter space consistent with a 125 GeV Higgs boson
including 3-loop corrections in the calculation of the Higgs
mass. These corrections increase $m_h$ by 1--3 GeV, lowering the
preferred scalar mass scale and decreasing the fine-tuning measure in
these scenarios. We then systematically examine prospects for dark
matter direct and indirect detection.  Direct detection constraints do
not exclude these models, especially for $\mu < 0$.  At the same time,
the scenario generically predicts spin-independent signals just beyond
current bounds.  We also consider indirect detection with neutrinos,
gamma rays, anti-protons, and anti-deuterons.  Current IceCube
neutrino constraints are competitive with direct detection, implying
bright prospects for complementary searches with both direct and
indirect detection.
\end{abstract}



\maketitle

\section{Introduction}

There are now many experimental constraints on weak-scale
supersymmetry.  These exclude generic supersymmetric theories in which
all superpartners have masses below a TeV, and focus attention on the
remaining supersymmetric theories that are both phenomenologically
viable and natural. In this work, we consider focus point
supersymmetry~\cite{Feng:1999mn,Feng:1999zg}, in which multi-TeV
squarks and sleptons are hierarchically heavier than the other
superpartners.

Focus point models are motivated by a variety of considerations.
Heavy first and second generation sfermions help satisfy low-energy
constraints on flavor and CP violation, and heavy third generation
sfermions raise the Higgs boson mass to the required level of 125
GeV~\cite{:2012gk,:2012gu}.  There are also theoretical reasons for
expecting scalar superpartners to be heavier than the gauginos.  For
example, such a hierarchy results from an approximate U(1)$_R$
symmetry~\cite{Feng:1999zg} or if none of the supersymmetry-breaking
fields is a complete gauge
singlet~\cite{Randall:1998uk,Giudice:1998xp}.  Note also that gaugino
masses enter the scalar mass renormalization group (RG) equations, but
scalar masses do not enter the gaugino mass RG equations; letting
$M_{1/2}$ and $m_0$ denote generic gaugino and scalar masses,
respectively, the hierarchy $m_0 \gg M_{1/2}$ is therefore stable
under RG evolution, whereas $M_{1/2} \gg m_0$ is not.  Last, although
large supersymmetry-breaking parameters are generically associated
with significant fine-tuning of the Higgs potential, simple
correlations in high-scale scalar mass parameters may reduce the
sensitivity of the weak scale to variations in these parameters,
providing a naturalness motivation for such models.

In this work, we consider in detail prospects for dark matter
detection in such theories~\cite{Feng:2000gh,Feng:2000zu}.  For
concreteness, we consider the focus point region of minimal
supergravity (mSUGRA), but the results are far more general: when the
scalar superpartners are very heavy, they effectively decouple from
dark matter phenomenology, and the details of the multi-TeV spectrum
are largely irrelevant.  The phenomenology of focus point dark matter
encompasses the phenomenology of mixed Bino-Higgsino and pure Higgsino
neutralino dark matter, and our conclusions for dark matter detection
are generically valid for any model with heavy scalars where the Bino
soft-supersymmetry breaking mass is lower than the Wino mass.

In \secref{parameter}, we explain our treatment of mSUGRA parameter
space.  We then turn to the Higgs mass in \secref{higgs}.  There have
been many studies of mSUGRA after the Higgs discovery; see, \eg,
Refs.~\cite{Akula:2012kk,Buchmueller:2012hv,Baer:2012mv,Strege:2012bt}.
In contrast to these, here we include a 3-loop calculation of the
Higgs mass using the public
code~\hthreem~\cite{Harlander:2008ju,Kant:2010tf}.  We find that
3-loop contributions raise the Higgs mass by 1--3 GeV over 2-loop
results.  Given the logarithmic sensitivity of the Higgs mass to the
top squark mass, this lowers the preferred range of stop masses
considerably.  In this calculation stop masses as low as 3 TeV are
consistent with the measured Higgs mass, even without significant stop
left-right mixing. In the focus point parameter space, this correlates
with a gluino as light as 2 TeV.

We then consider prospects for dark matter detection in the region of
parameter space preferred by the Higgs mass and other phenomenological
constraints, including direct searches for supersymmetric particles.
In \secref{dmdd}, we discuss both spin-independent and spin-dependent
direct detection and show that, contrary to claims in the literature,
perfectly viable regions of parameter space remain, especially for
$\mu < 0$.  Crucial to this conclusion is the small value for the
strange quark content of the nucleon now preferred by both lattice
calculations and chiral perturbation theory results.  At the same
time, the scenario generically predicts spin-independent cross
sections $\sigmaSI \sim 1~\zb = 10^{-9}~\pb = 10^{-45}~\cm^2$,
implying that dark matter candidates in this class of theories might
very well be discovered by direct detection experiments in the near
future.

In Secs.~\ref{sec:nu}, \ref{sec:gammas}, and \ref{sec:antimatter}, we
analyze the implications for indirect detection with neutrinos, gamma
rays, and anti-matter, respectively.  Although gamma rays and
anti-matter are currently not very constraining in focus point
scenarios, current bounds from observations of neutrinos from the
direction of the Sun with IceCube are stringent, and future runs with
planned upgrades will probe much of the preferred region, providing an
exciting, and in many respects orthogonal, complement to direct
detection.  In \secref{concl}, we discuss our results and conclude.

\section{Parameter Space and LHC Superpartner Searches}
\label{sec:parameter}

The defining feature of focus point supersymmetry is the insensitivity
of the weak scale to variations in the fundamental
supersymmetry-breaking parameters, even in the presence of multi-TeV
soft supersymmetry-breaking parameters.  Focus point supersymmetry
accommodates a range of thermal relic neutralinos that vary
continuously from $\sim 100~\gev$ Bino-Higgsino mixtures to heavier
and more Higgsino-like neutralinos, culminating in Higgsino-like
neutralinos with masses around 1 TeV~\cite{Feng:2000gh,Mizuta:1992qp}.
Given the appeal of neutralino dark matter, it is natural to impose
the thermal relic density as a constraint on the parameters space.  In
the context of mSUGRA, this constraint allows for a departure from the
typical $(m_0, M_{1/2})$ parameter space --- in which the
cosmologically viable region is only a small sliver --- to a parameter
space in which every point is cosmologically viable and more
parameters can be examined~\cite{Feng:2010ef}.  This parameter space
is particularly relevant in light of the first three years of LHC
results, which have effectively eliminated the so-called ``bulk''
scenario for neutralino dark matter with light scalars and severely
constrained coannihilation scenarios with light scalars, while leaving
the focus point relatively unscathed and strong as a possibility for
neutralino dark matter.

In mSUGRA, the relic density constraint can be cast as the requirement
that
\begin{equation}
\Omega_\chi \left(m_0, M_{1/2}, A_0, \tan\beta, 
\text{sign}(\mu) \right) = \Omega_{\text{DM}} \ ,
\label{omega}
\end{equation}
where $\Omega_{\text{DM}} \simeq
0.23$~\cite{Komatsu:2010fb,Ade:2013zuv} is the dark matter density in
units of the critical density.  Focus point supersymmetry is possible
with large $A$-parameters~\cite{Feng:2012jfa}, but given the
motivations of simplicity, the hierarchy between
supersymmetry-breaking parameters enforced by an approximate $U(1)_R$
symmetry, and the prediction of suppressed $A$-terms in some
high-energy frameworks~\cite{Dudas:2012wi,Evans:2013lpa}, we choose
$A_0 = 0$ throughout.  We may then use \eqnref{omega} to solve for
$m_0$ and present results in the $(\tan\beta, M_{1/2})$ plane for both
signs of $\mu$, with every point in these planes having the correct
relic density.  In general, \eqnref{omega} may be satisfied by more
than one value of $m_0$; for example, there may be a coannihilation
solution at low $m_0$ and a focus point solution at larger $m_0$.  In
such cases, we always use the largest allowed value of $m_0$.

\begin{figure}[tb]
\subfigure[ \ $\mu < 0$]{
\includegraphics[width=0.48\textwidth]{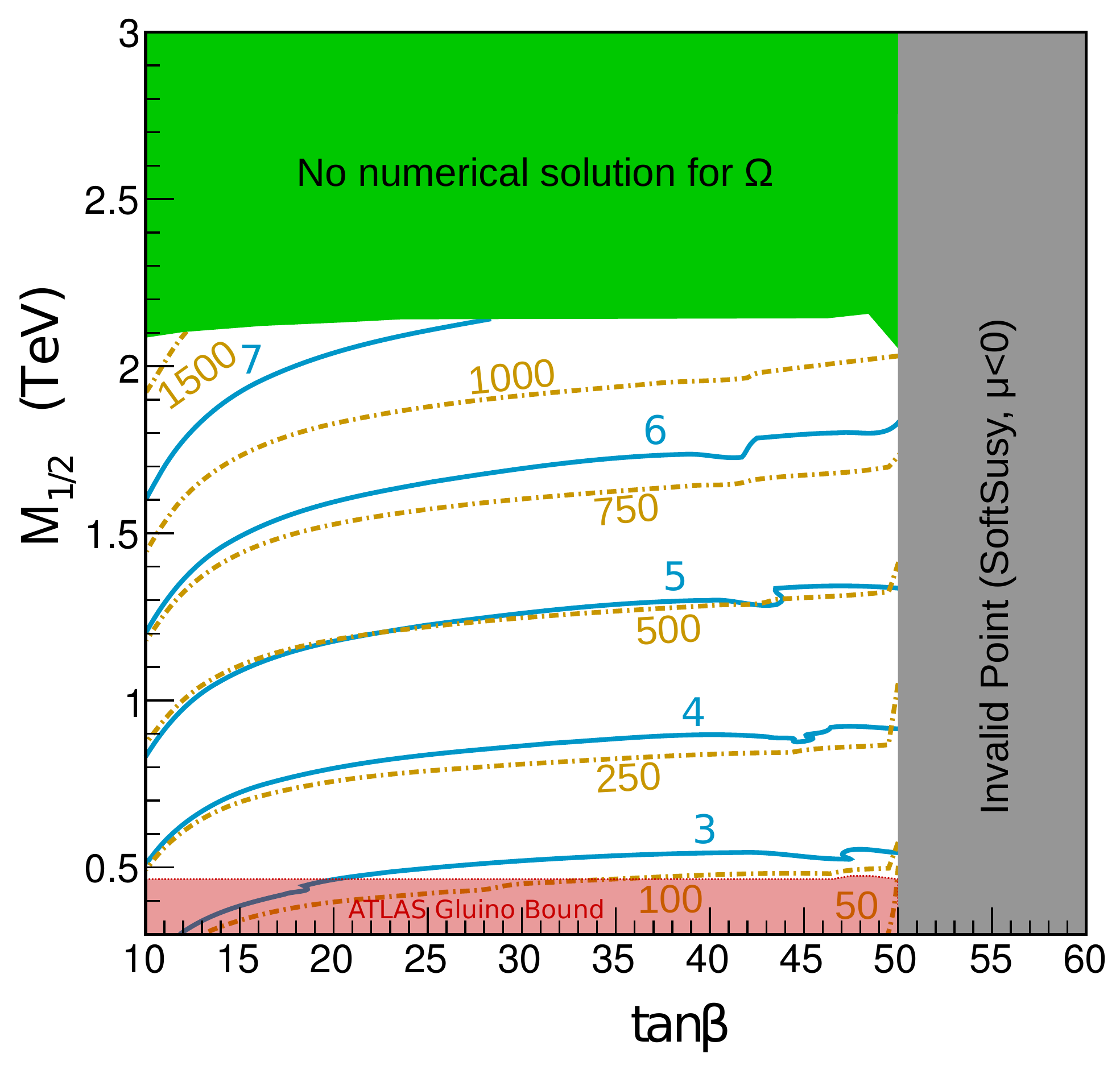}
\label{fig:munegm0finetune}
}
\subfigure[ \ $\mu > 0$]{
\includegraphics[width=0.48\textwidth]{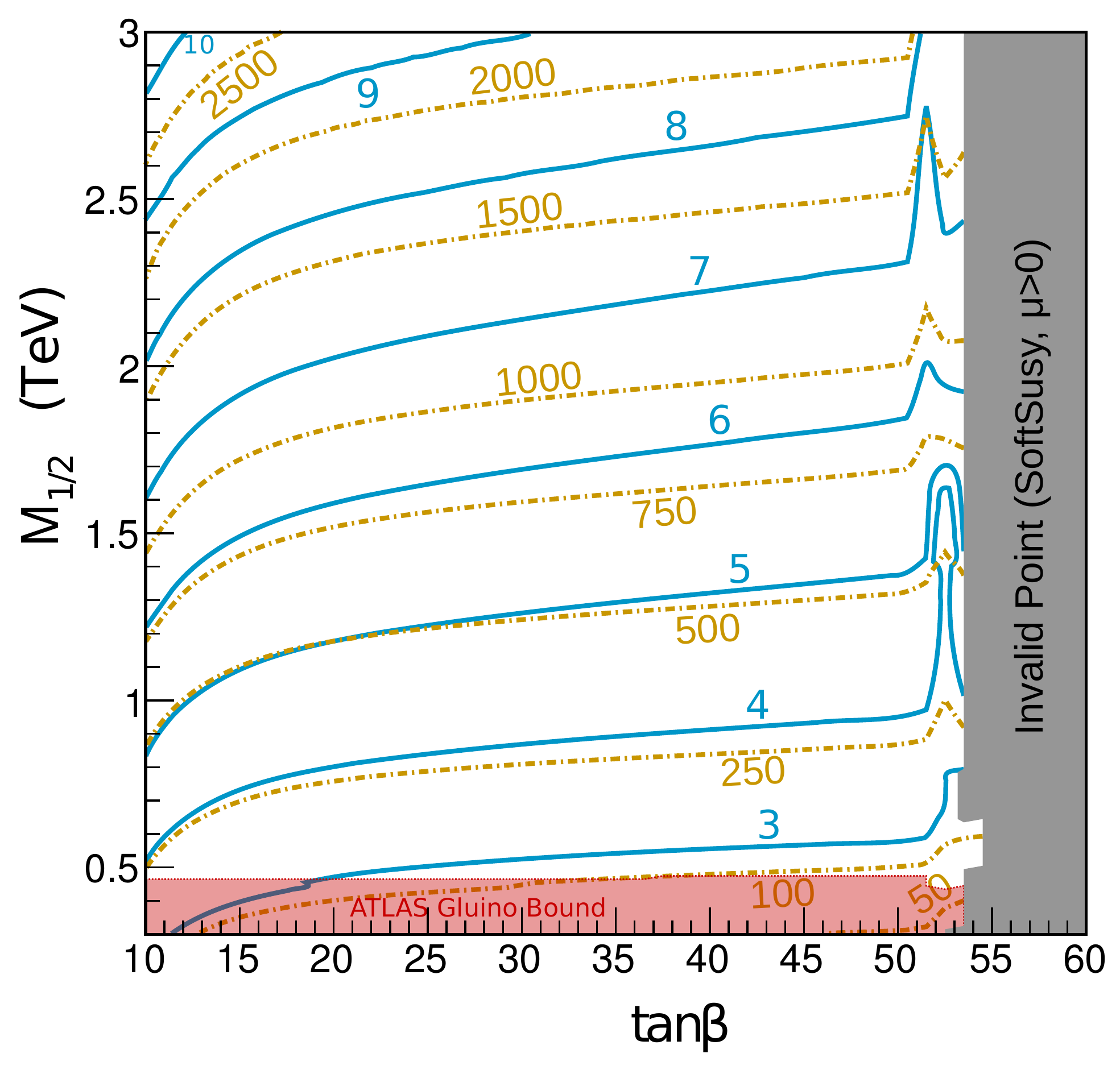}
\label{fig:muposm0finetune}
}
\caption{Contours of $m_0$ in TeV (solid blue) and fine-tuning
  parameter $c$ (dot-dashed gold) in the $(\tan\beta, M_{1/2})$ plane
  for $\Omega_\chi \simeq 0.23$, $A_0 = 0$ and $\mu < 0$ (left) and
  $\mu > 0$ (right).  The red shaded regions at low $M_{1/2}$ are
  excluded by the ATLAS gluino bound~\cite{ATLAS:2012ona}.  In the
  gray shaded regions at large $\tan\beta$, the RG evolution in
  SOFTSUSY becomes unreliable, and in the green shaded region at large
  $M_{1/2}$ for $\mu < 0$, numerical issues with loop corrections to
  neutralino masses make the solution algorithm for $\Omega$
  unreliable. }
\label{fig:m0finetune}
\end{figure}

\Figref{m0finetune} shows contours of $m_0$ in the $(\tan\beta,
M_{1/2})$ for points satisfying the relic density constraint, using
SOFTSUSY 3.1.7~\cite{Allanach:2001kg} to generate the SUSY spectrum
and MicrOMEGAs~2.4~\cite{Belanger:2010gh} to calculate the relic
density.  These solutions for $m_0$ are found for low values of
$|\mu|$ located near the $\mu^2 < 0$ region, where radiative
electroweak symmetry breaking fails.  The $\mu^2 < 0$ region moves to
higher $m_0$ for increasing $M_{1/2}$ and decreasing $\tan\beta$ due
to RG effects, and this behavior is reflected in the $m_0$ contours.
In \figref{m0finetune} the shaded region with low $M_{1/2}$ is
excluded by ATLAS searches for jets + missing
energy~\cite{ATLAS:2012ona}.  The other shaded regions, which will
appear in all of our figures, include a region at large $\tan\beta$,
where the RG evolution in SOFTSUSY becomes unreliable, and a region at
large $M_{1/2}$ for $\mu < 0$, where numerical issues with loop
corrections to neutralino masses make the solution algorithm for
$\Omega$ unreliable.  We stress that these last two regions are
excluded not by theoretical or experimental constraints, but rather
because numerical complications hinder our ability to make accurate
predictions.

\begin{figure}[tb]
\subfigure[ \ $\mu < 0$]{
\includegraphics[width=0.48\textwidth]{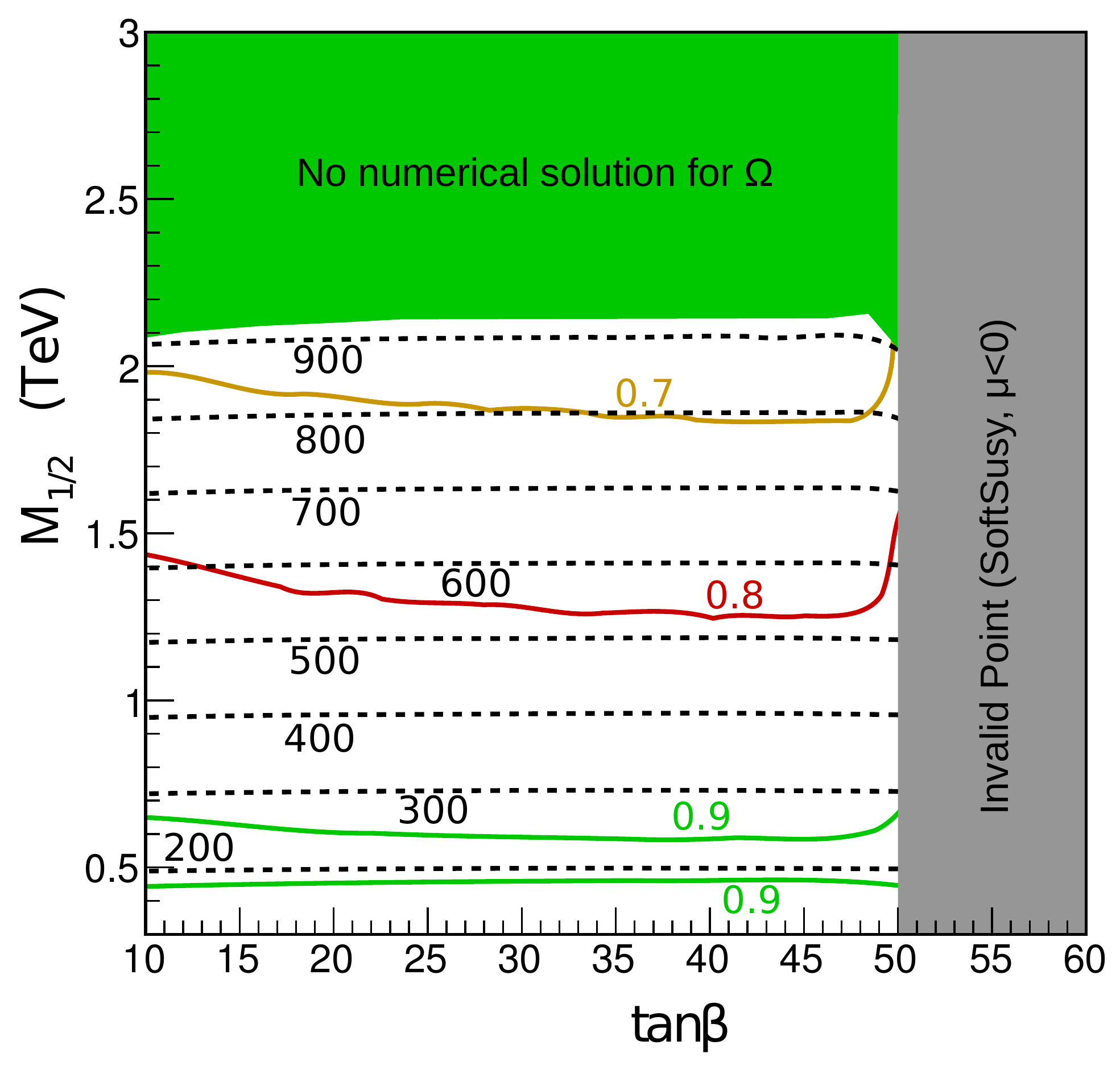}
\label{fig:munegmchiN11}
}
\subfigure[ \ $\mu > 0$]{
\includegraphics[width=0.48\textwidth]{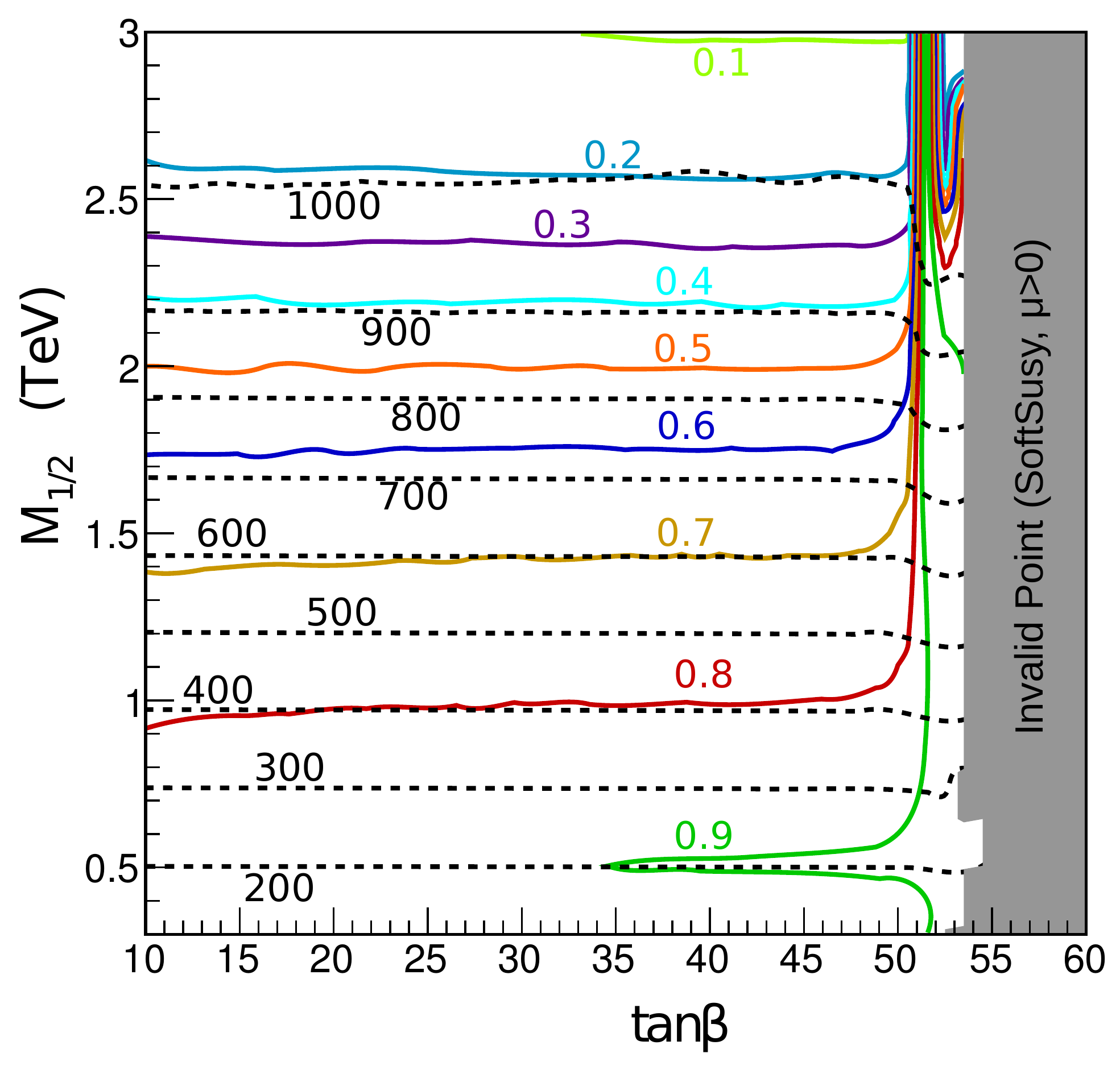}
\label{fig:muposmchiN11}
}
\caption{Contours of $m_\chi$ in GeV (black dotted) and
  $|a_{\tilde{B}}|$ (solid colored).}
\label{fig:mchiN11}
\end{figure}

Since the sfermion sector is decoupled in focus point supersymmetry,
the properties of neutralino dark matter are determined primarily by
its mass and the amount of Bino-Higgsino mixing present.  If the gauge
eigenstate composition of the lightest neutralino is given by
\begin{equation}
\chi = a_{\tilde{B}} (-i \tilde{B}) + a_{\tilde{W}} (-i \tilde{W}) +
a_{\tilde{H}_d} \tilde{H}_d + a_{\tilde{H}_u} \tilde{H}_u\ ,
\end{equation}
with $a_{\tilde{W}} \ll 1$ in the focus point region, the dominant
processes for both annihilation and scattering are proportional to
either $(a_{\tilde{B}} a_{\tilde{H}_{u,d}})^2$ or
$(a_{\tilde{H}_{u,d}})^4$~\cite{Feng:2010ef}.  Since
$|a_{\tilde{H}_u}| \sim |a_{\tilde{H}_d}|$, the mixing can be usefully
parameterized by the Bino content $a_{\tilde{B}}$.  \Figref{mchiN11}
contains contours of $m_\chi$ and $a_{\tilde{B}}$ consistent with
$\Omega_\chi = \Omega_{\text{DM}}$.  For much of the parameter space,
the neutralino dark matter is a Bino-Higgsino mixture, but as
$M_{1/2}$ increases, $m_\chi$ increases, and $a_{\tilde{B}}$
decreases: the increasing Higgsino content compensates for the
suppression of the annihilation cross-section by larger neutralino
masses to keep the thermal relic density constant.  The behavior is
similar for both signs of $\mu$, though $a_{\tilde{B}}$ is somewhat
larger in the $\mu < 0$ case relative to the $\mu > 0$ due to the
relative signs of $a_{\tilde{H}_{u,d}}$ for different signs of $\mu$.
In the limit of large $M_{1/2}$, the neutralino becomes nearly pure
Higgsino with $a_{\tilde{B}} \rightarrow 0$, and the neutralino mass
reaches $m_{\chi} \approx 1~\tev$.

In focus point scenarios, the weak scale is relatively insensitive to
variations in supersymmetry breaking parameters, allowing for
improved naturalness even with multi-TeV sfermion masses.  There are
many prescriptions for quantifying this naturalness, all of which are
subject to significant subjective choices; for a review, see
Ref.~\cite{Feng:2013pwa}.  Here we use a naturalness measure based on
the sensitivity coefficients~\cite{Ellis:1986yg,Barbieri:1987fn}
\begin{equation}
c_a \equiv \left| \frac{\partial \ln m_Z^2}{\partial \ln a^2}\right| \ ,
\label{ca}
\end{equation}
where $a^2$ is one of the input GUT-scale parameters $m_0^2$,
$\mgaugino^2$, $A_0^2$, $\mu_0^2$, and $m_3^2$, the $H^0_u H^0_d$ soft
mass parameter.  The overall fine-tuning of a model is defined as
\begin{equation}
c \equiv \text{max}\{c_a \} \ ,
\end{equation}
and contours of $c$ are shown in \figref{m0finetune}.  In the explored
region, $c_{m_0}$ is always the largest sensitivity coefficient, and
contours of $c$ roughly follow contours of $m_0$, with values of $m_0
\sim 4~\tev$ corresponding roughly to $c \sim 250$.  A subset of the
mSUGRA boundary conditions implies focusing, and the values of $c$
shown in \figref{m0finetune} are much smaller than would be expected
without the focus point behavior.

\section{Constraints from the Higgs Mass}
\label{sec:higgs}

\begin{figure}[tb]
\subfigure[ \ $\mu < 0$]{
\includegraphics[width=0.48\textwidth]{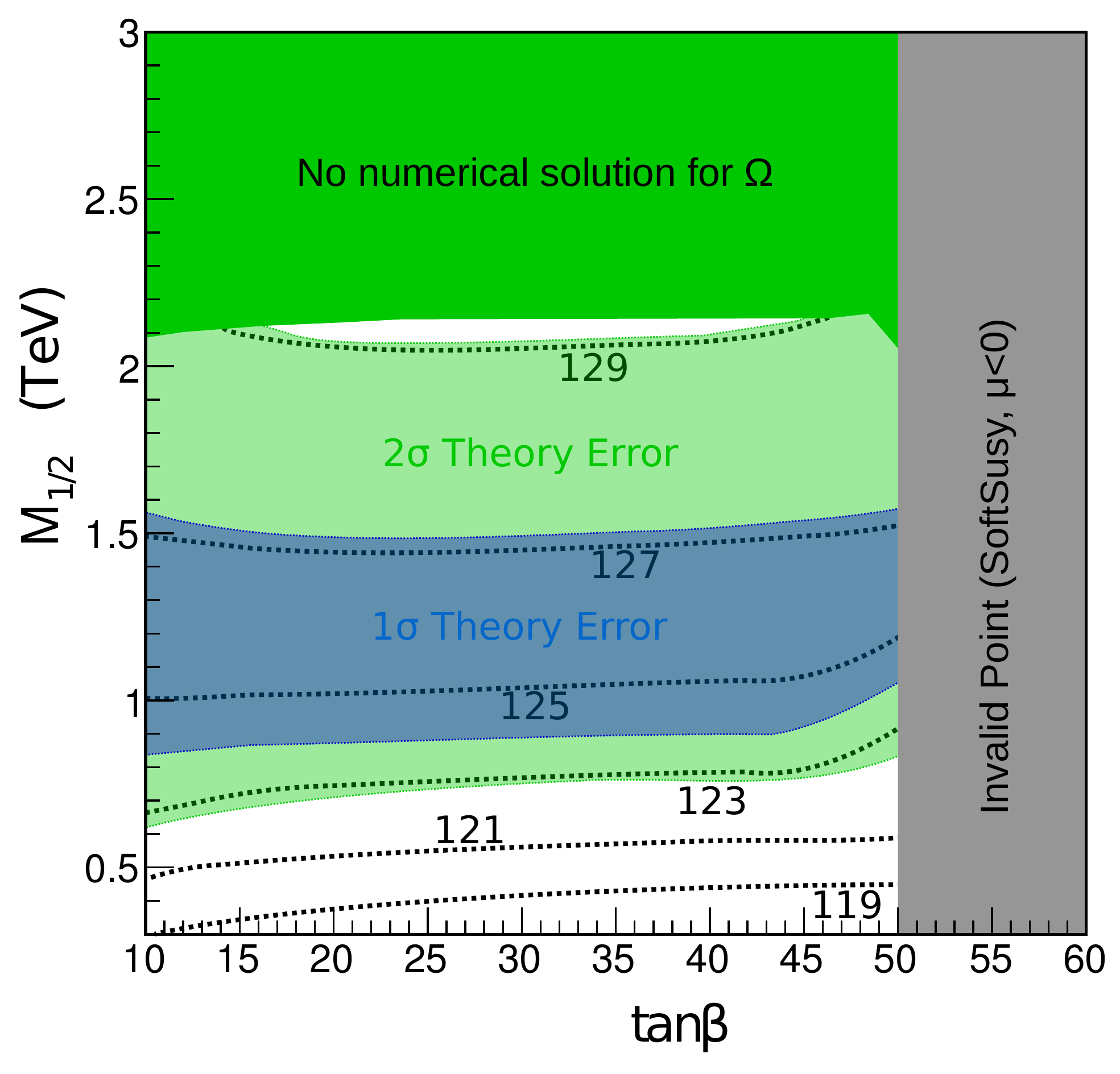}
\label{fig:munegmhiggs}
}
\subfigure[ \ $\mu > 0$]{
\includegraphics[width=0.48\textwidth]{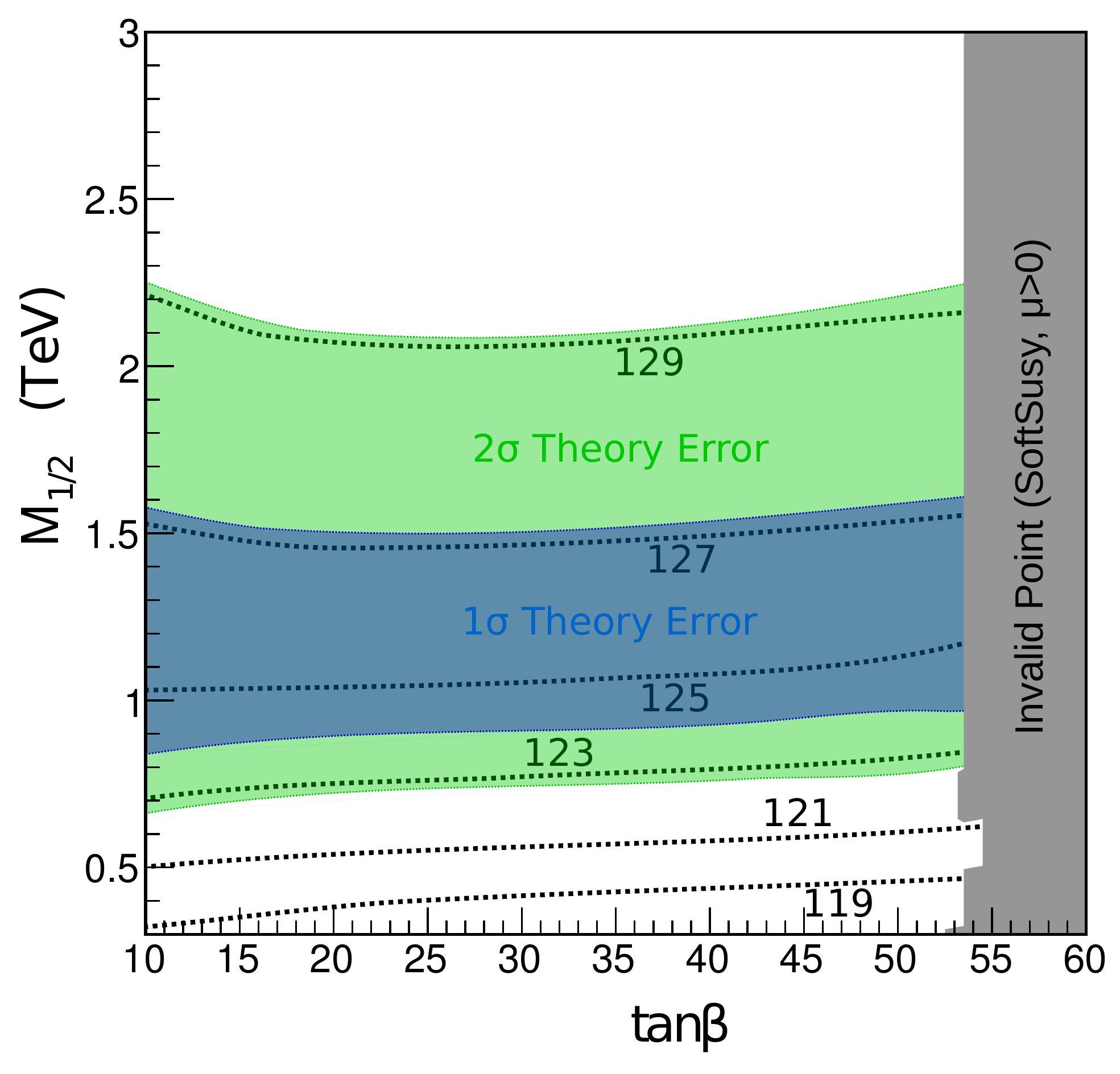}
\label{fig:muposmhiggs}
}
\caption{Contours of $m_h$ in GeV. In the shaded regions, the
  theoretical prediction for $m_h$ is within $1\sigma$ and
  $2\sigma$ of the experimental central value $m_h = 125.5~\gev$,
  where $\sigma^2 \equiv \Delta_{\text{th}}^2+(1~\gev)^2$.}
\label{fig:mh}
\end{figure}

The mass of the recently-discovered SM-like Higgs
boson~\cite{:2012gk,:2012gu} provides a stringent constraint on the
parameter space of any supersymmetric model. The most recent mass
measurements are~\cite{ATLASMoriond2013,CMSMoriond2013}
\begin{eqnarray}
\text{ATLAS $4\ell$}: &&         
124.3{~}^{+0.6}_{-0.5}{~}^{+0.5}_{-0.3}~\gev \\
\text{ATLAS $\gamma \gamma$}: && 126.8 \pm 0.2 \pm 0.7~\gev \\
\text{CMS $4\ell$}: &&           125.8 \pm 0.5 \pm 0.2~\gev \\
\text{CMS $\gamma \gamma$}: &&   125.4 \pm 0.5 \pm 0.6 ~\gev \ ,
\end{eqnarray}
where the first uncertainties are statistical and the second
uncertainties are systematic.

We calculate the lightest Higgs mass in the focus point region of
mSUGRA with the program \hthreem, which calculates $m_h$ in the
\drbar~scheme including the dominant 3-loop contributions at
$\mathcal{O}(\alpha_t
\alpha_s^2)$~\cite{Harlander:2008ju,Kant:2010tf}. In addition, we
modified~\hthreem~to increase the precision in the calculation of the
running~\drbar~top quark mass.\footnote{These changes are incorporated
  in the current version of~\hthreem, which has been released
  simultaneously with Ref.~\cite{Feng:2013tvd}.} We set
$m_t^{\text{pole}}=173.2$~GeV and $\alpha_s(m_Z)=0.1184$, and fix the
renormalization scale to the geometric mean of the stop masses.  For
further details, see Ref.~\cite{Feng:2013tvd}.

In \figref{mh}, we plot contours of $m_h$ in the parameter space
defined by \figref{m0finetune}.  We find that the 3-loop terms
generate a 1--3 GeV increase in $m_h$ over the 2-loop truncation.  The
2-loop terms in turn generate a 5--8 GeV increase over the 1-loop
truncation, indicating convergence of the series. We observe also that
the improved treatment of $m_t^{\text{\drbar}}$ and
$\alpha_s^{\text{\drbar}}$~increases the 2-loop prediction relative to
\FeynHiggs~\cite{Frank:2006yh,Degrassi:2002fi,Heinemeyer:1998np,%
  Heinemeyer:1998yj}.  For comparison, note that the geometric mean of
the stop masses ranges from about 1 TeV at low $M_{1/2}$ to 8 TeV at
high $M_{1/2}$ in the plotted parameter space.

In \figref{mh} we shade regions where the difference between the
calculated $m_h$ and the tentative central value 125.5 GeV is within
the indicated theoretical uncertainty in the calculation.  At each
point on the parameter space, we assign a theoretical error bar
$\Delta_{\text{th}}$, defined as
\begin{align}
&\Delta_{\text{th}}  \equiv \sqrt{(\Delta_{\text{pert}})^2 
+ (\Delta_{\text{para}})^2} \ , \nonumber\\
& \Delta_{\text{pert}}\equiv\frac{1}{2} \left| m_h^{\text{(3-loop)}} 
- m_h^{\text{(2-loop)}} \right| \;,\nonumber\\
& \Delta_{\text{para}} \equiv m_h(m_t=174.2~\gev)-m_h(m_t=173.2~\gev) \ .
\end{align}
The uncertainty $\Delta_{\text{pert}}$ from higher-order terms in the
perturbation series is estimated to be in the range 0.5--1.5
GeV.\footnote{The size of the 3-loop corrections is consistent within
  the uncertainty with the next-to-leading logarithm analysis of
  Ref.~\cite{Martin:2007pg}, which used a somewhat different
  organization of the perturbation series.}  The parametric
uncertainty $\Delta_{\text{para}}$ induced by the uncertainty in the
top quark mass is typically of order 0.5--1 GeV in the focus point
parameter space.

The positive 3-loop terms significantly impact the preferred range of
superpartner masses.  Requiring that the theoretical prediction be
within $\sqrt{\Delta_ {\text{th}}^2+(1~\gev)^2}$ of 125.5 GeV (where we
  have included a representative experimental uncertainty of 1 GeV
  based on the difference between $ZZ$ and $\gamma\gamma$ channels at
  ATLAS), scalar mass parameters as low as $m_0 \sim 4~\tev$,
  corresponding to stop masses as low as 3 TeV, and gluino masses as
  low as $m_{\tilde{g}} \approx 2.8 M_{1/2} \sim 2~\tev$ are
  consistent with the measured Higgs mass. Note that, combining the
  results shown in \figsref{m0finetune}{mh}, the 3-loop $m_h$
  contributions also decrease allowed values of the fine-tuning
  parameter $c$ by a factor of $\sim 5$.

\section{Direct Dark Matter Detection}
\label{sec:dmdd}

It is well-known that thermally-produced neutralinos can possess a
wide range of direct detection cross sections, from those that are
significantly excluded to those that are orders of magnitude below
current sensitivities.  However, this full range of cross sections is
not generic.  Highly suppressed direct detection is typically
associated with pure Bino scenarios, which have the correct thermal
relic density only if there are light sfermions, co-annihilation, or
resonant annihilation through the pseudoscalar Higgs resonance.  The
first two possibilities are disfavored by the non-observation of light
squarks at the LHC, while the third depends upon careful tuning of the
pseudoscalar Higgs mass to $m_A \approx 2 m_\chi$.  Most of the
remaining parameter space is populated by models with Bino-Higgsino
mixing like that found in the focus point region.  For these models,
the Bino-Higgsino mixing also sets the spin-independent
neutralino-proton scattering cross section, which falls in the range
$\sigmaSI \sim 1 - 40~\zb$ for a wide range of model parameters when
neutralinos have the right thermal relic density~\cite{Feng:2010ef}.
This range of cross sections is particularly relevant for current and
near-future direct detection experiments; the XENON100
experiment~\cite{Aprile:2011hi,Aprile:2012nq} has begun probing this
range of relevant cross sections, and near-future direct detection
experiments will be sensitive to most of the focus point region of
mSUGRA.

In the focus point region, $\sigmaSI$ is dominated by Higgs-mediated
diagrams, and the Higgs-neutralino coupling is sensitively dependent
on the sign of $\mu$, producing a suppression of $\sigmaSI$ in the
$\mu < 0$ case relative to the $\mu > 0$ case.  For moderate
$\tan\beta$ this leads to a relative factor of a few in $\sigmaSI$,
from the coupling coefficients and at large $\tan\beta$ due to the
relative contribution of the heavy Higgs-mediated diagrams.  Although
the general lore holds that $\mu > 0$ is preferable to address the
discrepancy in the anomalous magnetic moment of the
muon~\cite{Bennett:2006fi,Davier:2010nc,Jegerlehner:2011ti}, in focus
point theories the contribution for either sign of $\mu$ is too small
to produce consistency or to further aggravate the discrepancy without
considering significant non-universality of smuon
masses~\cite{Feng:2011aa}.

Determinations of $\sigmaSI$ for neutralinos also suffer from the
well-known uncertainty in the quark scalar form factor of the
nucleons, $f_q^N$, defined as
\begin{equation}
\left\langle N \left| m_q \bar{\psi}_q \psi_q \right| N \right\rangle
= f_q^N M_N \ .
\end{equation}
The form factors for the up- and down-type quarks are well measured,
and the heavy quark form factors are determined by loop contributions
from the gluon form factor, but there is a longstanding controversy
regarding the strange quark form factor, which feeds into $\sigmaSI$
in a quantitatively important
way~\cite{Ellis:2008hf,Giedt:2009mr,Buchmueller:2011ki}.  Older
results from chiral perturbation
theory~\cite{Borasoy:1996bx,Gasser:1990ap,Bernard:1996nu} combined
with determination of the nucleon sigma term from meson scattering
data~\cite{Pavan:2001wz}, and supported by direct
computation~\cite{Alarcon:2011zs}, suggested $f_s = f_s^n = f_s^p
\approx 0.36$.  For this value of $f_s$, the other form factors are
all much smaller, $f_{q \neq s}^N \alt 0.05$, and so the strange quark
contribution dominates the direct detection cross
section~\cite{Ellis:2008hf}.  However, recent lattice studies favor a
much smaller value of~\cite{Young:2009zb,Freeman:2009pu}
\begin{equation}
f_s \approx 0.05 \ ,
\end{equation}
much closer to the other quark
flavors~\cite{Giedt:2009mr,Thomas:2011cg}.  It has also been argued
that the lower value for $f_s$ is consistent with chiral perturbation
theory computations, provided higher-order baryon decuplet
contributions are taken into
account~\cite{Jenkins:1991bs,Bernard:1993nj,Young:2009zb,Alarcon:2011zs}.
A recent calculation considering these contributions found $f_s =
0.017 \pm 0.15$~\cite{Alarcon:2012nr}; for similar recent conclusions,
see Refs.~\cite{Junnarkar:2013ac,Young:2013nn}.  Here we take $f_s =
0.05$ in deriving direct detection cross sections.

\begin{figure}[tb]
\subfigure[ \ $\mu < 0$]{
\includegraphics[width=0.48\textwidth]{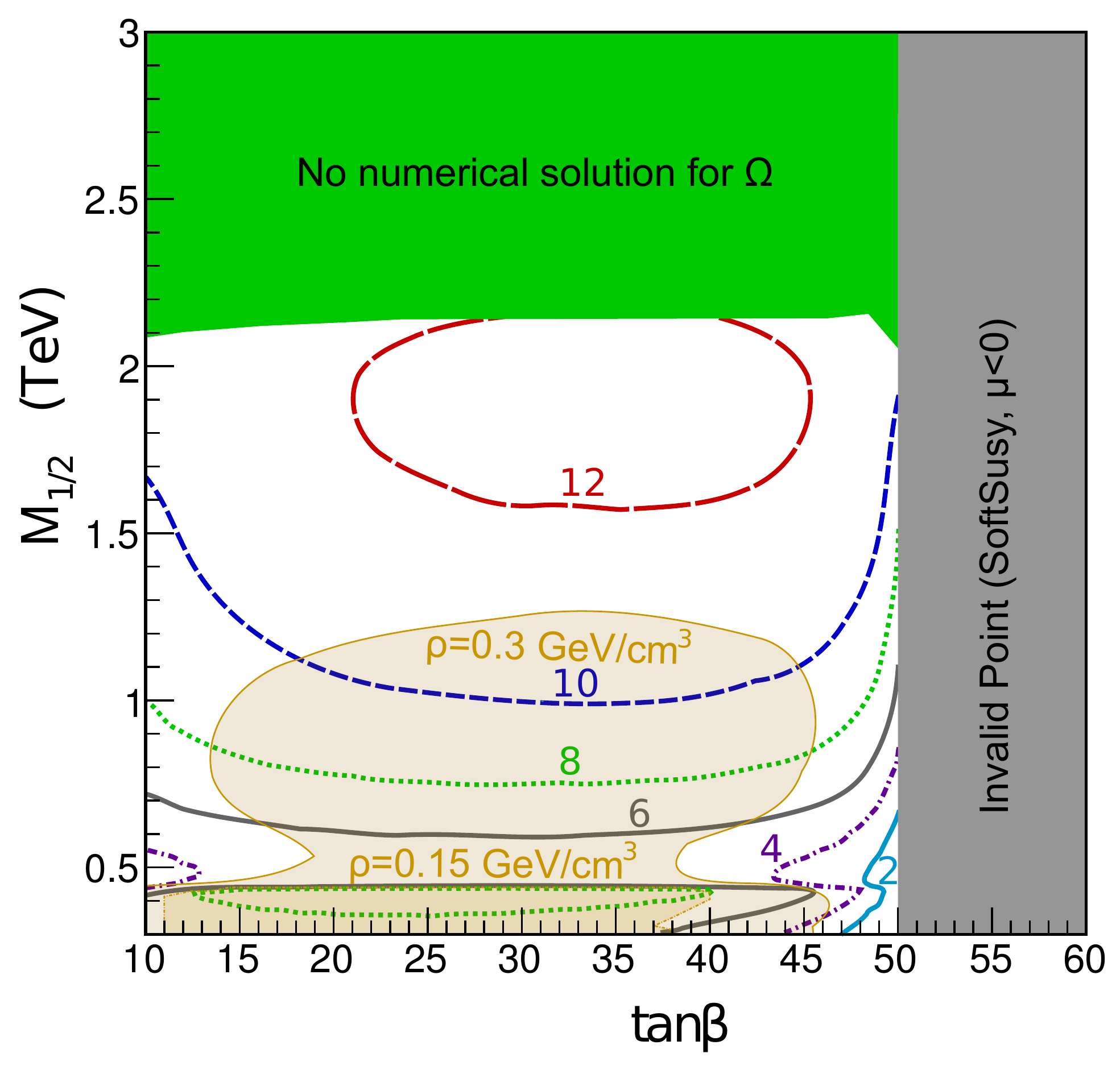}
\label{fig:munegddsi}
}
\subfigure[ \ $\mu > 0$]{
\includegraphics[width=0.48\textwidth]{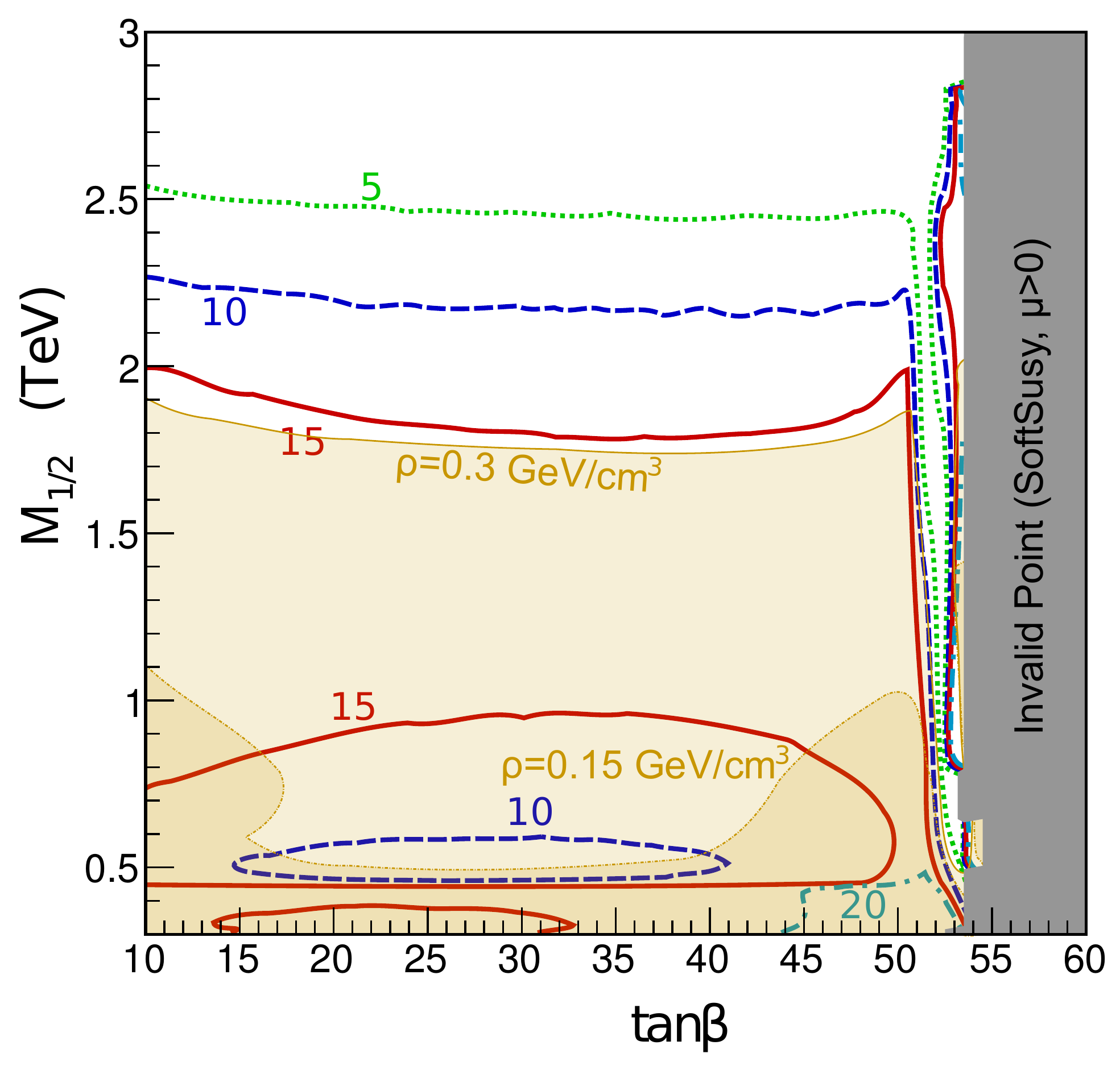}
\label{fig:muposddsi}
}
\caption{Contours of spin-independent scattering cross section
  $\sigmaSI$ in zb.  The shaded regions are excluded by
  XENON100~\cite{Aprile:2012nq}, assuming local dark matter density
  $\rho_{\text{local}} = 0.3~\gev/\cm^3$ (light shaded) and
  $\rho_{\text{local}} = 0.15~\gev/\cm^3$ (dark shaded).}
\label{fig:ddsi}
\end{figure}

\Figref{ddsi} shows exclusion contours for XENON100 in the
$(\tan\beta, M_{1/2})$ plane for both signs of $\mu$ and
$\rho_{\text{local}} = 0.3~\gev/\cm^3$.  For $\mu > 0$, current
XENON100 bounds require $M_{1/2} \agt 1.8~\tev$ for a wide range of
$\tan\beta$, with stronger exclusions at large and small $\tan\beta$,
as discussed above.  A small region at very large $\tan\beta$ is
allowed for $M_{1/2} \agt 500~\gev$; here the lightest neutralino is
nearly pure Bino due to the $A$-funnel crossing through the focus
point region.  For $\mu < 0$, XENON100 requires $M_{1/2} \agt
1.3~\tev$ for moderate values of $\tan\beta$, but the exclusions are
much weaker for small and large $\tan\beta$. For small $\tan\beta$,
this is because of suppression of the dark matter-Higgs coupling from
the interplay of the two Higgsino components, and at large
$\tan\beta$, it is caused by a cancelation between the light and heavy
Higgs diagrams~\cite{Feng:2011aa}.  As a result, large portions of the
parameter space remain viable.  Exclusion contours for
$\rho_{\text{local}} = 0.15~\gev/\cm^3$ are also presented, motivated
by the possibility of a local dark matter density somewhat lower than
normal due to the presence of small-scale
structure~\cite{Kamionkowski:2008vw}.  For this lower value of
$\rho_{\text{local}}$ and both signs of $\mu$, the excluded region is
roughly comparable to that excluded by gluino searches, only becoming
stronger for large and small $\tan\beta$ when $\mu > 0$, and almost
none of the parameter space preferred by the Higgs mass is excluded by
direct detection.

\begin{figure}[tb]
\subfigure[ \ $\mu < 0$]{
\includegraphics[width=0.48\textwidth]{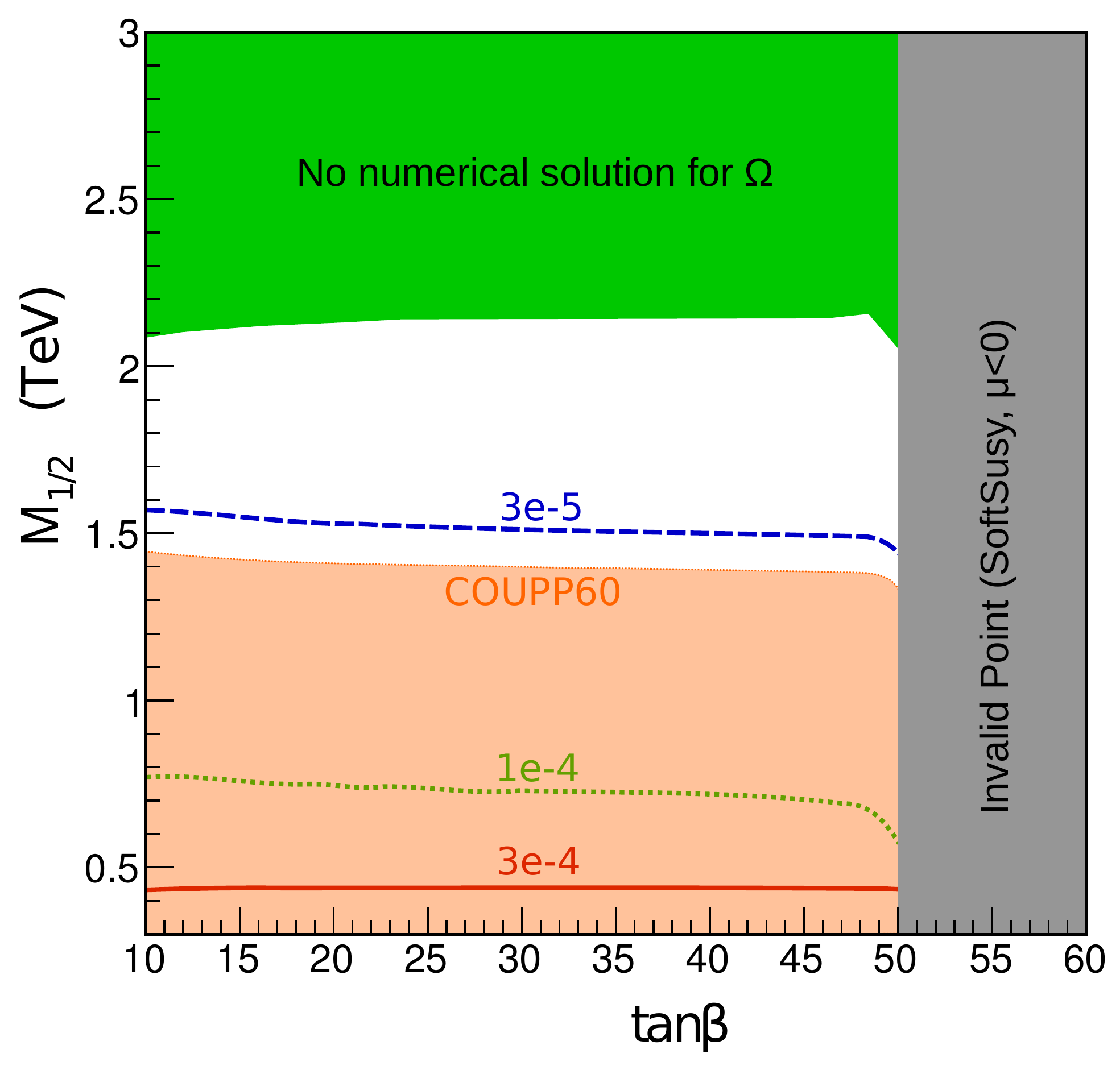}
\label{fig:munegddsd}
}
\subfigure[ \ $\mu > 0$]{
\includegraphics[width=0.48\textwidth]{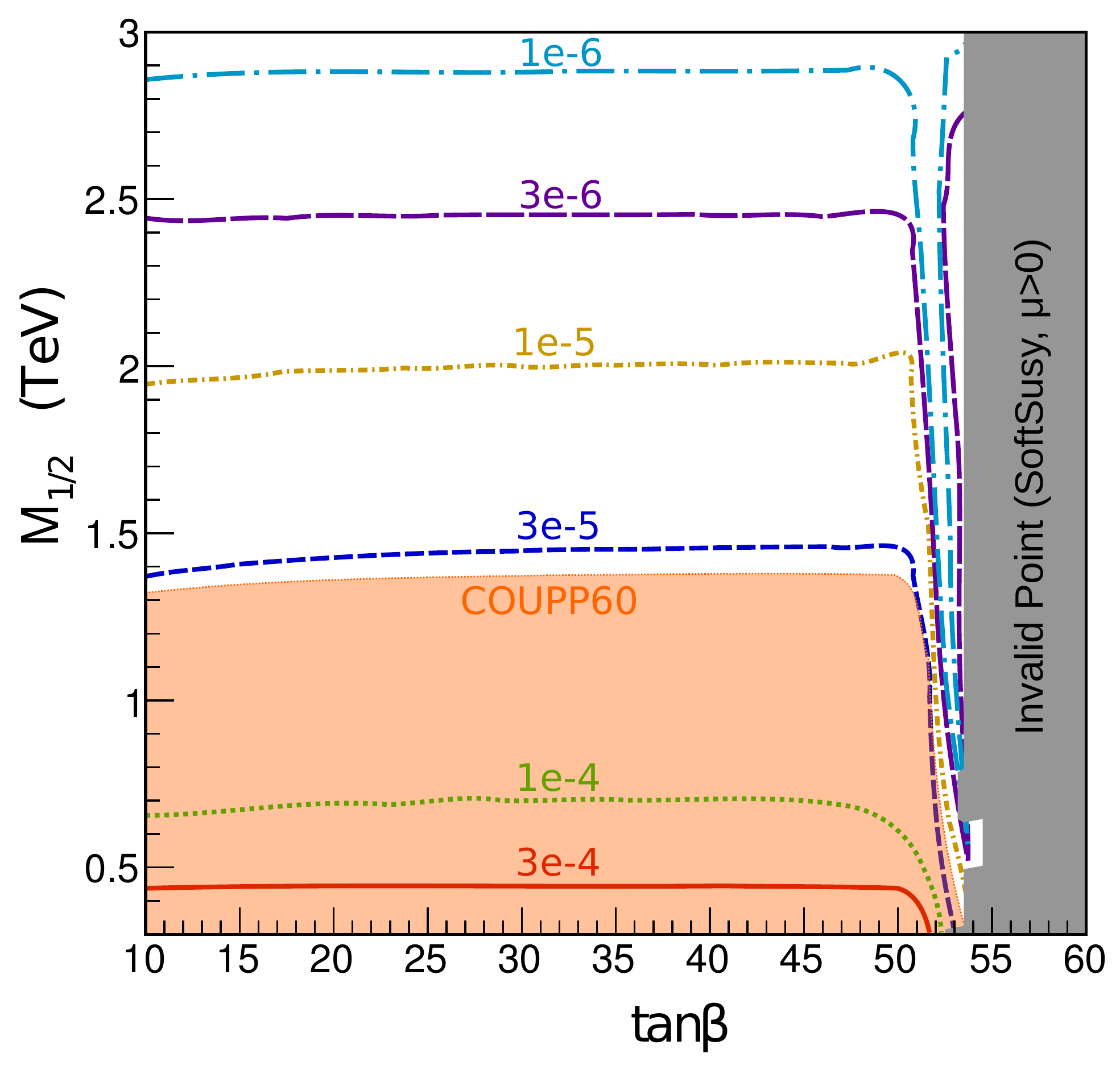}
\label{fig:muposddsd}
}
\caption{Contours of the spin-dependent neutralino-proton scattering
  cross section $\sigmaSD$ in pb. The shaded region indicates the
  reach of COUPP-60~\cite{coupp60} after 12 months with
  $\rho_{\text{local}} = 0.3~\gev/\cm^3$.}
\label{fig:SD}
\end{figure}

Dark matter may also be detected directly through its spin-dependent
couplings.  \Figref{SD} shows contours of constant $\sigmaSD$, the
spin-dependent neutralino-proton scattering cross section. Across the
parameter space of the focus point region compatible with the correct
thermal neutralino relic density, $\sigmaSD$ is in the range $10^{-6}
- 3\times10^{-4}$ pb, for both signs of $\mu$, decreasing with
increasing $M_{1/2}$.  At large values of $M_{1/2}$, the lightest
neutralino becomes increasingly Higgsino-like, suppressing $\sigmaSD$.
However, the observed Higgs mass disfavors the pure Higgsino limit,
and the $2\sigma$ allowed region for $m_h$ favors $\sigmaSD$ in the
range $10^{-4} - 10^{-5}$ pb.

The shaded region shows the sensitivity expected from
COUPP-60~\cite{coupp60,coupptalk}, corresponding to a data-taking
period of 12 months at SNOLAB, in the zero-background assumption and
using the typical local dark matter density of $\rho_{\text{local}} =
0.3~\gev/\cm^3$. With one year of data, the COUPP-500kg experimental
sensitivity is anticipated to range between a few $\times 10^{-6}~\pb$
at 100 GeV to a few $\times 10^{-5}~\pb$ at 1 TeV, thus covering a
significant portion of the parameter space of interest here.

\section{Neutrinos from Annihilation in the Sun and in the Earth}
\label{sec:nu}

The search for high-energy neutrinos from the direction of the center
of the Sun or of the Earth has a special place in the ranks of
indirect detection techniques. In the limit where the capture rate of
dark matter particles in celestial bodies is equilibrated by the
annihilation rate, the flux of neutrinos solely depends on the
scattering cross section of dark matter off of nuclei in the celestial
bodies. In the case of the Sun, the dominant scattering mechanism for
neutralinos in the minimal supersymmetric standard model is typically
spin-dependent scattering, while scattering in the Earth is dominated
by spin-independent processes. Unlike searches for antimatter or gamma
rays, where the target dark matter densities are generally poorly
known and affected by large uncertainties, the flux of neutrinos from
the Sun or the Earth has a rather mild dependence on astrophysical
inputs. The only crucial information is, in fact, the local dark
matter density. In this respect, of all indirect searches, neutrino
telescopes provide perhaps the most robust limits.

\begin{figure}[tb] 
\subfigure[ \ $\mu < 0$]{
\includegraphics[width=0.48\textwidth]{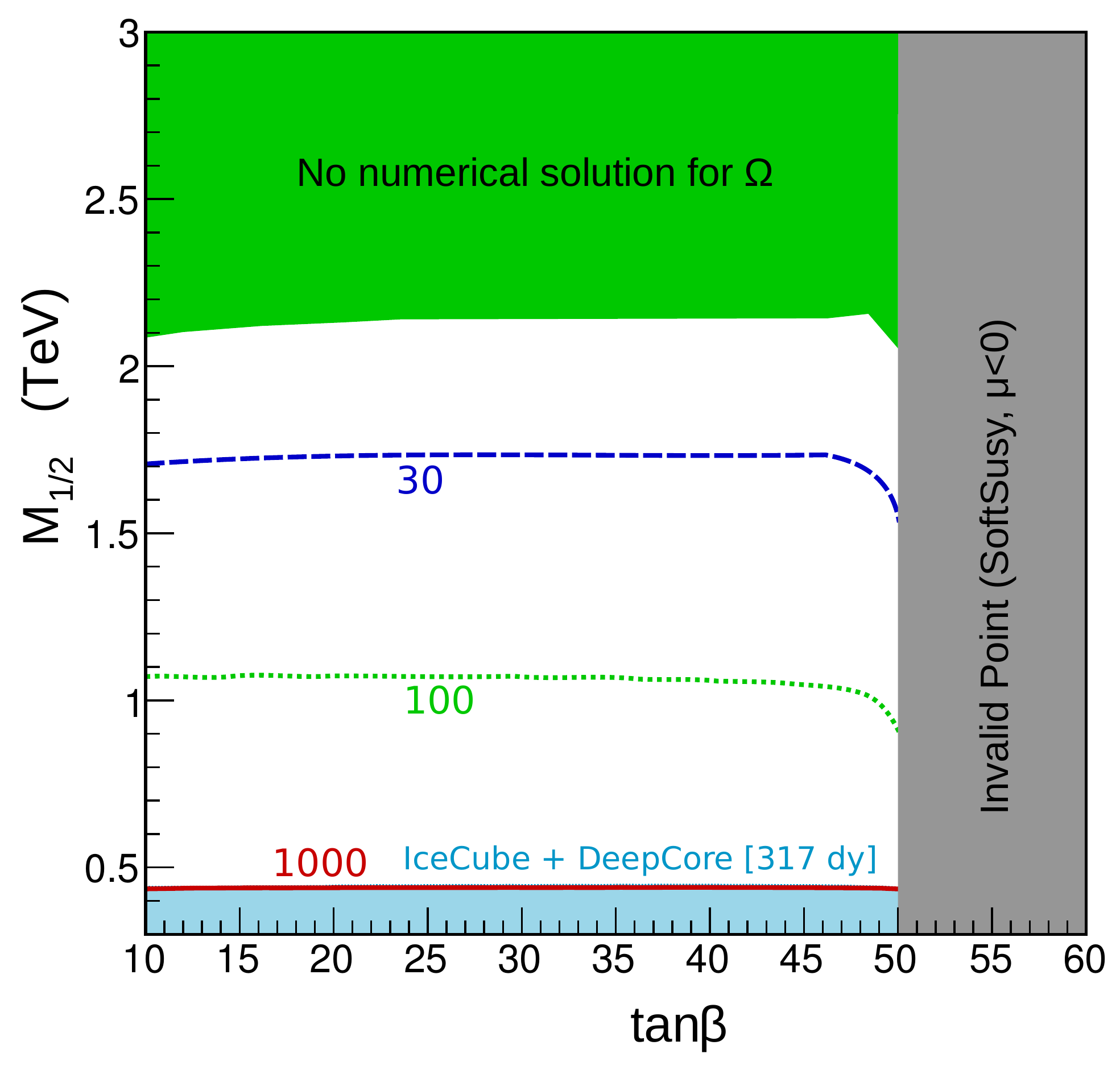}
\label{fig:munegnuSun1}
}
\subfigure[ \ $\mu > 0$]{
\includegraphics[width=0.48\textwidth]{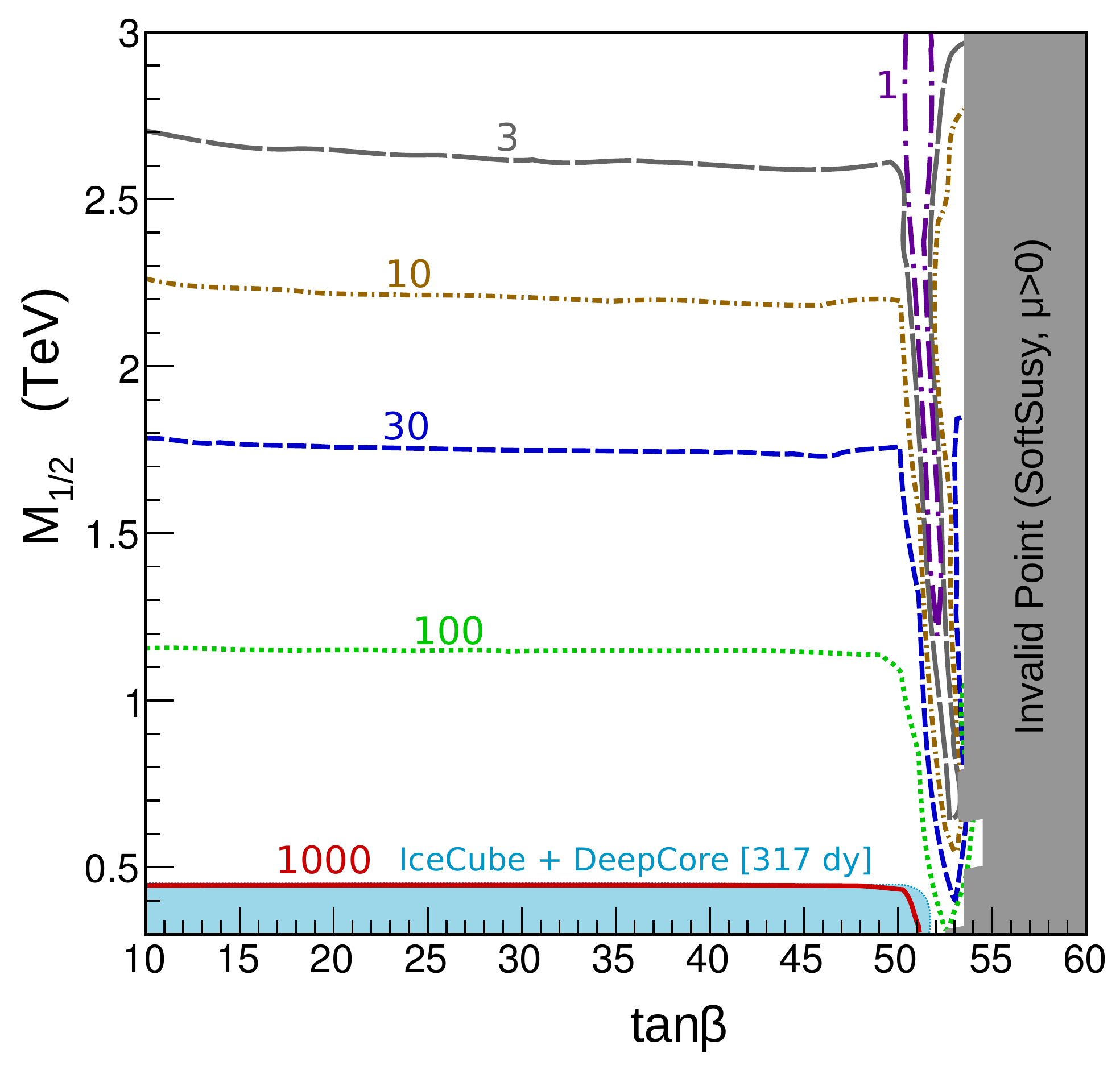}
\label{fig:muposnuSun1}
}
\caption{The flux of muons in units of $\km^{-2}~\yr^{-1}$ with
  energies above 1 GeV at IceCube.  The shaded region is excluded by
  current limits from IceCube/DeepCore~\cite{Aartsen:2012kia}.}
\label{fig:nuSun1}
\end{figure}

In \figref{nuSun1} we show the flux of muons produced via
charge-current interactions by high-energy neutrinos from dark matter
annihilation in the Sun. To calculate this rate (as well as all of the
subsequent indirect detection rates) we employ the DarkSUSY package,
version 5.0.5~\cite{ds}. \Figref{nuSun1} shows the integrated muon
rate for muons with energies larger than 1 GeV. The shaded region at
the bottom is excluded by the latest results from 317 days of data
taken from 2010-11 at the IceCube neutrino telescope with the
79-string configuration, and with the use of the DeepCore
sub-array~\cite{Aartsen:2012kia}.  This region excludes a parameter
space portion comparable to that excluded by current LHC
searches. Note that the 1 GeV threshold is much lower than the
detector's actual energy threshold, even with the use of DeepCore, but
the 1 GeV threshold is used in Ref.~\cite{Aartsen:2012kia} for
consistency with other results in the field, especially from
experiments such as SuperKamiokande, where the 1 GeV threshold is
actually experimentally meaningful.  For IceCube/DeepCore, the
extrapolation below the 1 GeV threshold is made based on the assumed
neutrino spectrum, which in the focus point region corresponds closely
to the $W^+W^-$ channel for which the exclusion limits are quoted in
Ref.~\cite{Aartsen:2012kia}.

\begin{figure}[tb]
\subfigure[ \ $\mu < 0$]{
\includegraphics[width=0.48\textwidth]{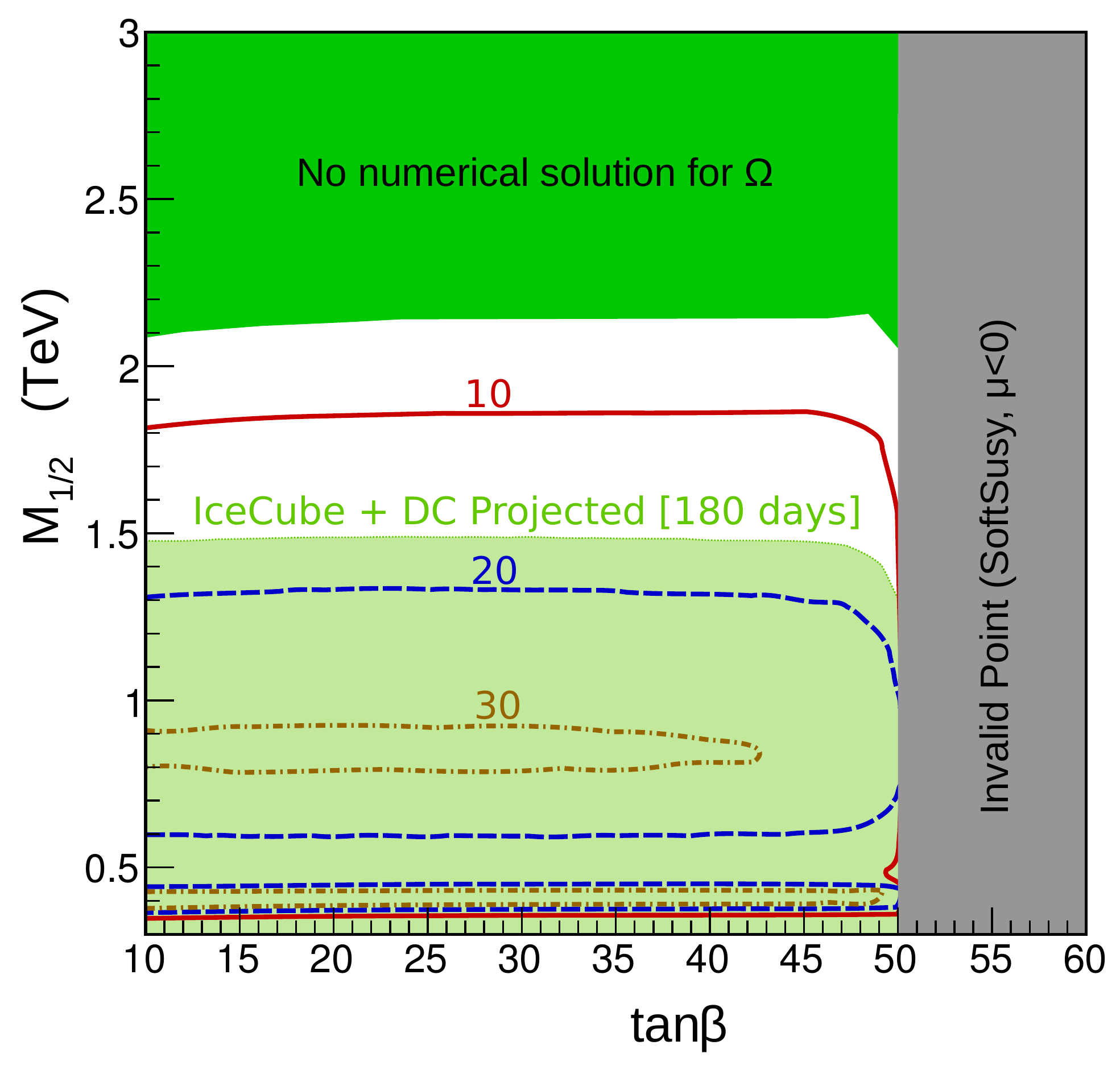}
\label{fig:munegnuSun100}
}
\subfigure[ \ $\mu > 0$]{
\includegraphics[width=0.48\textwidth]{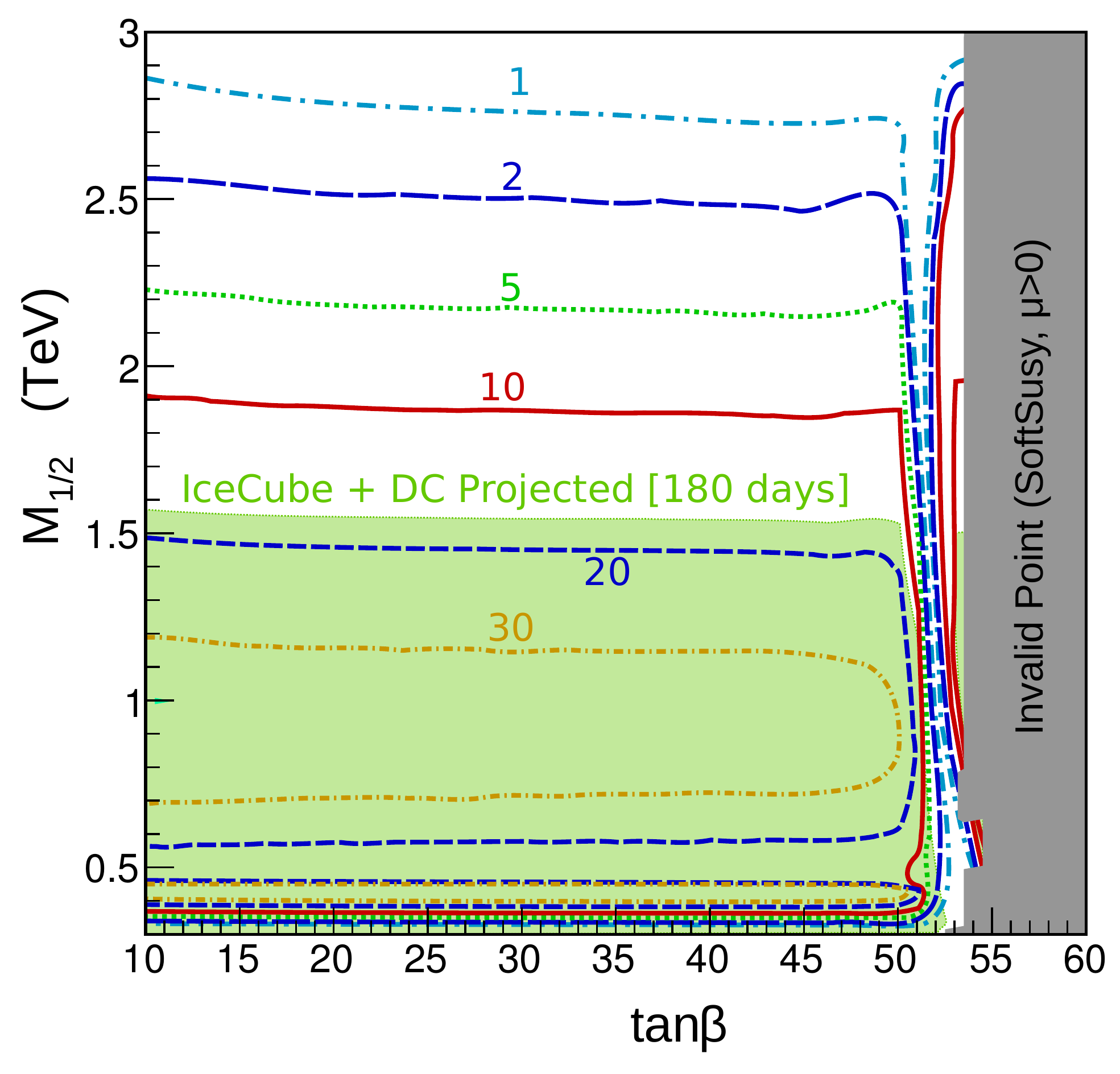}
\label{fig:muposnuSun100}
}
\caption{The flux of muons in units of $\km^{-2}~\yr^{-1}$ with
  energies above 100 GeV at IceCube.  The shaded region shows the
  originally anticipated sensitivity region for 180 live days for the 
  IceCube/DeepCore system~\cite{IceCube:2011aj}.}
\label{fig:nuSun100}
\end{figure}

The results of Ref.~\cite{Aartsen:2012kia} fell short by about a
factor 2--5 of the anticipated target sensitivity quoted in
Ref.~\cite{IceCube:2011aj} for 180 days. We find that had the detector
performed to the level anticipated in Ref.~\cite{IceCube:2011aj}, the
exclusion limit would have extended up to $M_{1/2} \approx 1.5~\tev$,
covering much of the parameter space of the focus point region
compatible with the Higgs mass.  This is supported by
\figref{nuSun100}, where we show the flux of muons integrated above a
100 GeV threshold; these numbers are therefore more indicative of the
actual number of events IceCube might detect than those shown by the
contours of \figref{nuSun1}.  The shaded region corresponds to the
original 180 days sensitivity target, which would have excluded
$M_{1/2} \alt 1.5~\tev$ with little dependence on $\tan\beta$,
corresponding to a lightest neutralino mass of $\sim 600~\gev$.  This
emphasizes how promising neutrino telescope searches are in the
context of searches for a signal of new physics from the focus point
region. We also note that the recent null results from the ANTARES
collaboration~\cite{Adrian-Martinez:2013ayv} reinforce the lack of a
high-energy neutrino signal from the Sun, at a level very close to the
current IceCube/DeepCore limits.

\begin{figure}[tb]
\subfigure[ \ $\mu < 0$]{
\includegraphics[width=0.48\textwidth]{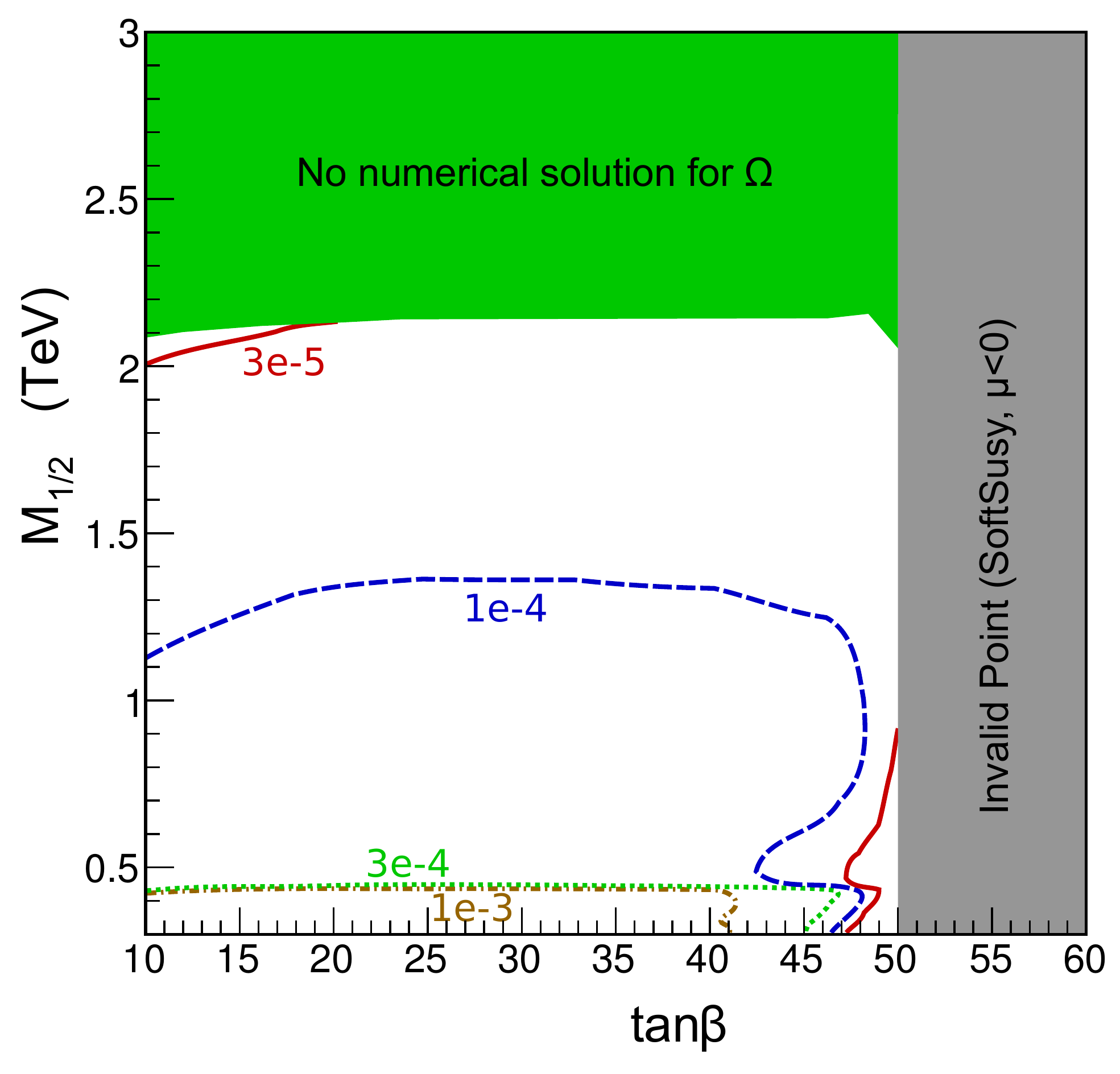}
\label{fig:munegnuEarth1}
}
\subfigure[ \ $\mu > 0$]{
\includegraphics[width=0.48\textwidth]{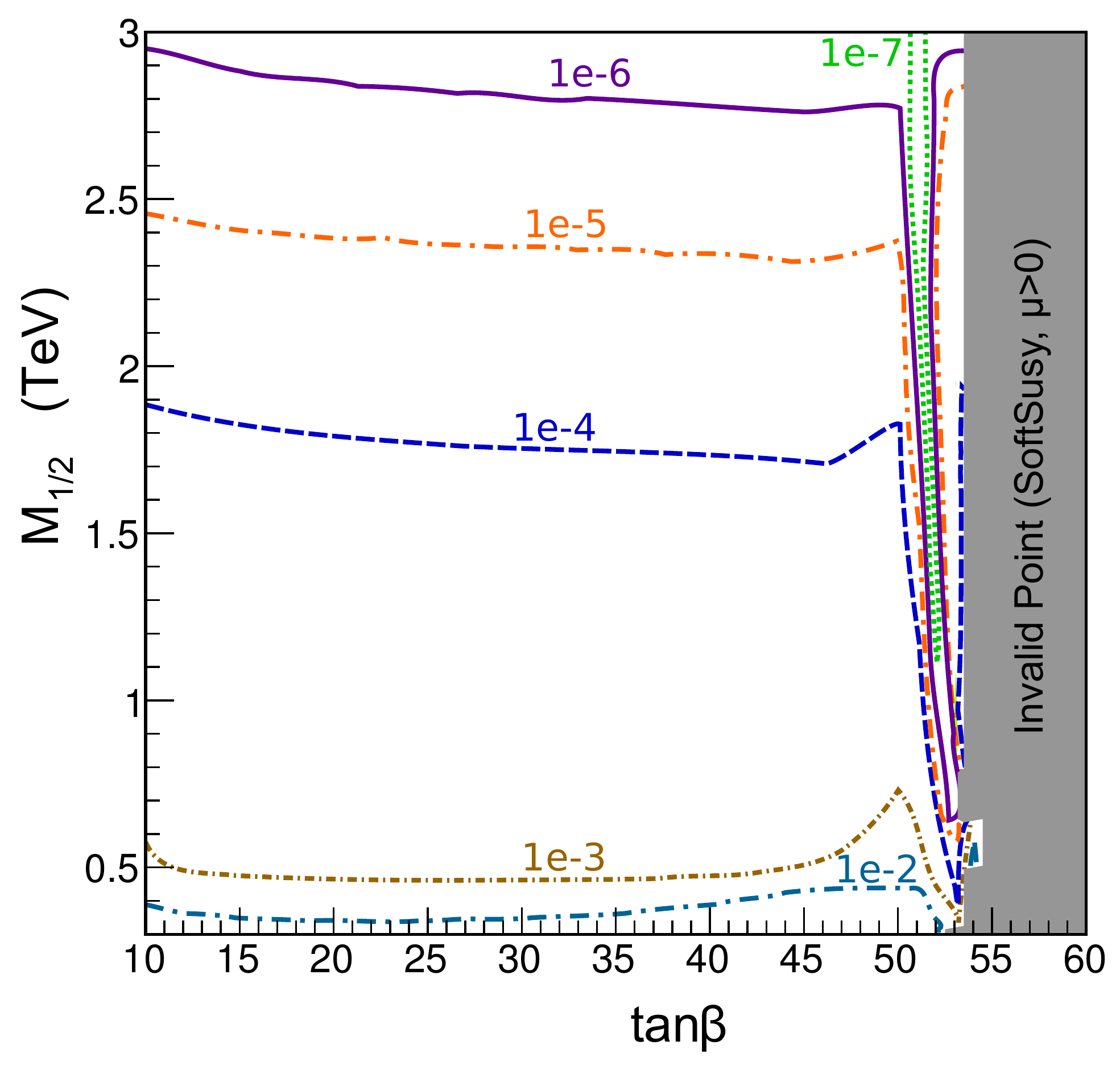}
\label{fig:muposnuEarth1}
}
\caption{The flux of muons in units of $\km^{-2}~\yr^{-1}$ with
  energies above 1 GeV at IceCube from dark matter annihilation in the
  center of the Earth.}
\label{fig:nuEarth1}
\end{figure}
The rates of high-energy neutrinos, and consequently of muons, from
neutralino annihilation in the center of the Earth are not nearly as
exciting as those from the center of the Sun.  We show in
\figref{nuEarth1} the calculated fluxes of muons from the Earth, again
integrated above a 1 GeV energy threshold.  Nowhere do we obtain
fluxes much larger than $10^{-3}$ km$^{-2}$ yr$^{-1}$, which is
clearly well below the sensitivity of km$^3$-sized neutrino
telescopes. We note that unlike the case of the Sun, for the Earth the
dependence of the flux of neutrinos on the spin-independent cross
section induces a significant dependence on the sign of $\mu$, with
positive $\mu$ producing larger fluxes due to the lack of interfering
terms in the neutralino-proton scalar cross section, as discussed in
the previous section.

\section{Gamma Rays}
\label{sec:gammas}

\begin{figure}[tb] 
\subfigure[ \ $\mu < 0$]{
\includegraphics[width=0.48\textwidth]{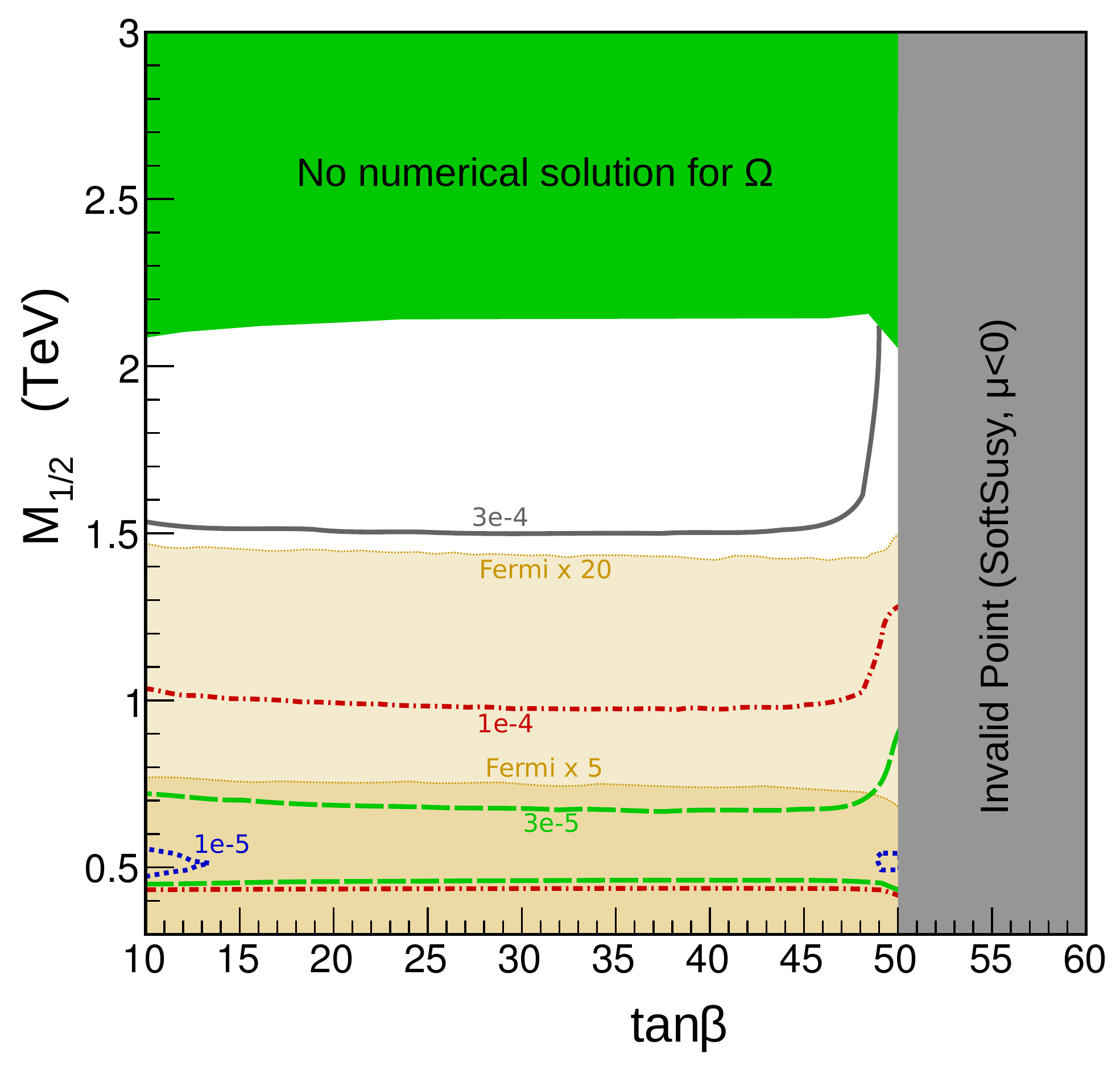}
\label{fig:munegbryy}
}
\subfigure[ \ $\mu > 0$]{
\includegraphics[width=0.48\textwidth]{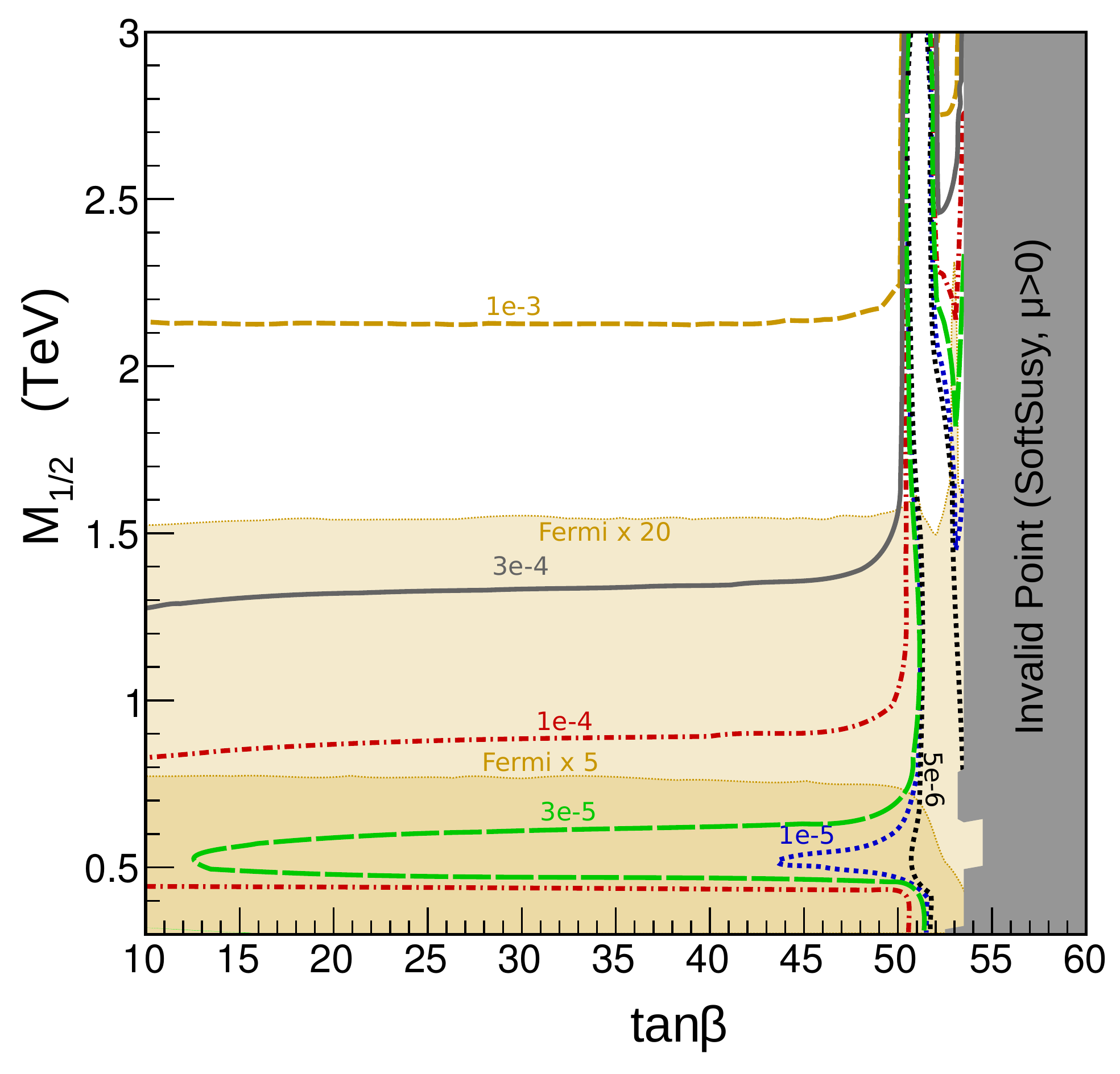}
\label{fig:muposbryy}
}
\caption{Gamma ray searches in the focus point region. The curves
  indicate constant values for the branching ratio for neutralino pair
  annihilation into two photons.  Null results from Fermi searches for
  a monochromatic gamma-ray line do not put any constraint on this
  plane. Null results for continuum gamma-ray signals with Fermi using
  stacked dwarf galaxies also do not exclude any of this parameter
  space.  However, improvements of current bounds on the gamma-ray
  continuum will probe the parameter space.  The shaded regions
  indicate the performance of searches for a continuum gamma-ray
  signal, assuming current sensitivities are improved by factors of 5
  and 20, as indicated.}
\label{fig:bryy}
\end{figure}

Gamma rays provide another promising possibility for the indirect
detection of dark matter.  This signal is especially relevant now that
the Fermi Large Area Telescope (LAT)~\cite{lat} has revolutionized our
understanding of the high-energy sky, in a photon energy range
extraordinarily relevant for indirect searches for WIMP dark matter.

The gamma-ray signal may take one of two forms.  It may appear as a
monochromatic line, if photons are produced as one or both of the
annihilation products in a two-body final state.  Alternatively, the
signal may be an excess of continuum gamma rays extending for several
decades in energy below the dark matter particle mass.  Such continuum
gamma rays are typically produced from the two-photon decay of neutral
pions resulting from the hadronization of annihilation products, or
from final state radiation, or from inverse Compton processes
associated with final state electrons and positrons.

We begin by considering the line signal.  \Figref{bryy} shows curves
of constant branching ratio into two photons.  The branching fraction
increases towards increasing masses,\footnote{This might be partly due
  to the fact that the annihilation mode into two photons is the only
  electroweak one-loop correction implemented in DarkSUSY: this
  artificially boosts the branching ratio into two photons as the
  neutralino mass approaches $M_W/\alpha_W$~\cite{nojiri}; this result
  should therefore be taken with a grain of salt.} but is always much
smaller than the percent level.  In the focus point parameter space,
the thermally-averaged neutralino pair annihilation cross section
always lies at about $1 - 2 \times 10^{-26}~\cm^3~\s^{-1}$, with the
mismatch with the canonical value of $3\times 10^{-26}~\cm^3~\s^{-1}$
being due to chargino and next-to-lightest neutralino
coannihilation. These values imply that the Fermi LAT Collaboration
line search limits~\cite{fermiline} do not yet constrain this
parameter space.

The recent discovery of a 130 GeV line-like feature in the Fermi LAT
data has attracted great attention~\cite{bringman,weniger}.  Our
results indicate that focus point supersymmetry does not provide a
viable framework to explain the line feature with dark matter
annihilation, as the branching ratio into two photons, and the
associated pair-annihilation cross section, are much smaller than the
required value of $\sim 10^{-27}~\cm^3~\s^{-1}$.

Turning next to the continuum signals, we consider annihilation in
local dwarf galaxies, currently one of the most stringent and robust
limits on the pair-annihilation cross section of dark matter.  Cross
sections of the order of what the theory predicts over the parameter
space of interest are only constrained for neutralino masses on the
order of 30 GeV~\cite{fermidsph}. In focus point supersymmetry, such
masses are never consistent with the relic density constraint (and are
also excluded by neutrino telescope searches and by LHC results), and
the limits weaken approximately quadratically with mass.

This is illustrated with the shaded regions shown in \figref{bryy},
which indicate the improvement to the Fermi limits needed to probe the
parameter space of interest; we indicate the sensitivity lines
corresponding to improvements by factors of 5 and 20.  In the focus
point region, neutralinos pair-annihilate with a branching ratio close
to 100\% into SU(2) gauge boson pairs, $WW$ and $ZZ$. The two channels
produce very similar gamma-ray spectra. To determine the limits from
the Fermi combined dwarf observations, we therefore employed the $WW$
final state limits shown in that work.  To approach the level of
$M_{1/2}\sim 1.5~\tev$, the Fermi limit from stacked dwarf
galaxies~\cite{fermidsph} would need to be improved by a factor of
20. Such an improvement would take a time-frame which is beyond the
anticipated lifetime of the mission.  We note, however, that an
improvement of a factor 5 corresponds approximately to observations of
the same 10 dSph employed in the current Fermi LAT limits, but for an
observation time of 10 years~\cite{Morselli:2013wqa}.

As presented in \figref{bryy}, the constraints from gamma-ray
observations are notably less effective than those from neutrino
telescopes.  A comparison between the two methods is not trivial: in
all models under consideration here there exists equilibration between
neutralino capture and annihilation in the Sun. The neutrino flux from
the Sun thus depends almost exclusively on the capture rate which, in
turn, depends on the spin dependent scattering cross section. This is
an entirely different quantity from the ratio of annihilation rate
over neutralino mass squared that enters the Fermi constraints. The
large energy threshold for Neutrino Telescopes also affects the limits
in the low-mass region, while no such threshold effect is present for
the Fermi limits.

It is important to note, however, that we have considered here line
and continuum signals given conservative assumptions.  Constraints can
be obtained by employing optimistic choices for the density profile of
the inner Galaxy, or of external galaxies or clusters, or by utilizing
optimistic assuptions for the dark matter sub-structure content and
structure. Here, we have limited ourselves to the more conservative
limits obtained by the Fermi Collaboration for line
signals~\cite{fermiline} and continuum signals from stacked
dwarfs~\cite{fermidsph}. We emphasize that had we used the Galactic
center and a favorable dark matter density profile, we could have
easily reached radically more optimistic conclusions.

We do not show here predictions for the performance of a future
Cherenkov Telescope Array (CTA); see, e.g.,
Ref.~\cite{Doro:2012xx}. Certain sensitivity estimates for the reach
of CTA optimistically carve into the parameter space of the focus
point region, for example from observations of the inner
Galaxy~\cite{Doro:2012xx}. Interestingly, CTA will be especially
sensitive to WIMP masses in the TeV region, and is thus, in principle,
an ideal instrument to look for a signal in the focus point region.
Under conservative assumptions, however, CTA, like Fermi, is not
guaranteed to detect a signal from dark matter models in the focus
point region. In addition, annihilation of a 1 TeV neutralino in the
focus point region to the level needed for a detection with CTA would
lead to significant low-energy inverse Compton gamma-ray production,
which might conflict with existing Fermi LAT limits. We postpone
detailed discussion to future work, but we emphasize that CTA will be
a key observational tool in the search for particle dark matter in
this region, especially if a signal for TeV-mass dark matter were
detected in direct detection or neutrino telescope experiments.

\section{Antimatter}
\label{sec:antimatter}

The successful deployment of the Alpha Magnetic Spectrometer (AMS-02)
on board the International Space Station has boosted hopes and
expectations of using cosmic-ray antimatter as a probe of annihilation
of Galactic dark matter. In the context of the focus point region, for
models with the correct thermal neutralino relic density, the flux of
positrons is always too small to be detectable with any significance
by current experiments, so we focus here on anti-protons and
anti-deuterons. The latter choice is motivated by the extremely
suppressed background rate and great discrimination capabilities
against anti-protons that the future General Antiparticle Spectrometer
(GAPS) mission promises for anti-deuterons in the low energy
(approximately at or below 1 GeV)
range~\cite{Hailey:2013gwa,Baer:2005tw}.

\begin{figure}[tb]
\subfigure[ \ $\mu < 0$]{
\includegraphics[width=0.48\textwidth]{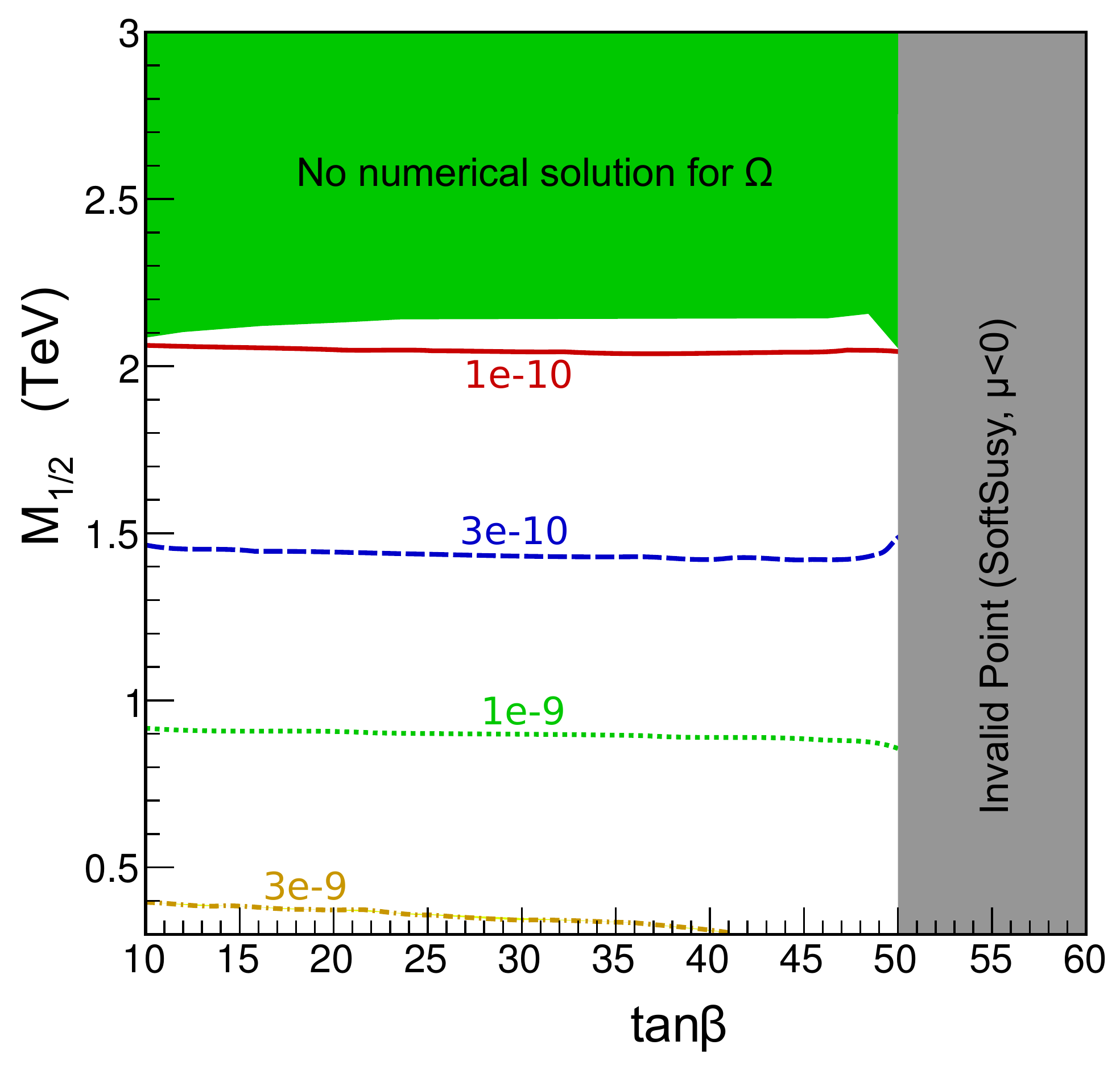}
\label{fig:munegpbar}
}
\subfigure[ \ $\mu > 0$]{
\includegraphics[width=0.48\textwidth]{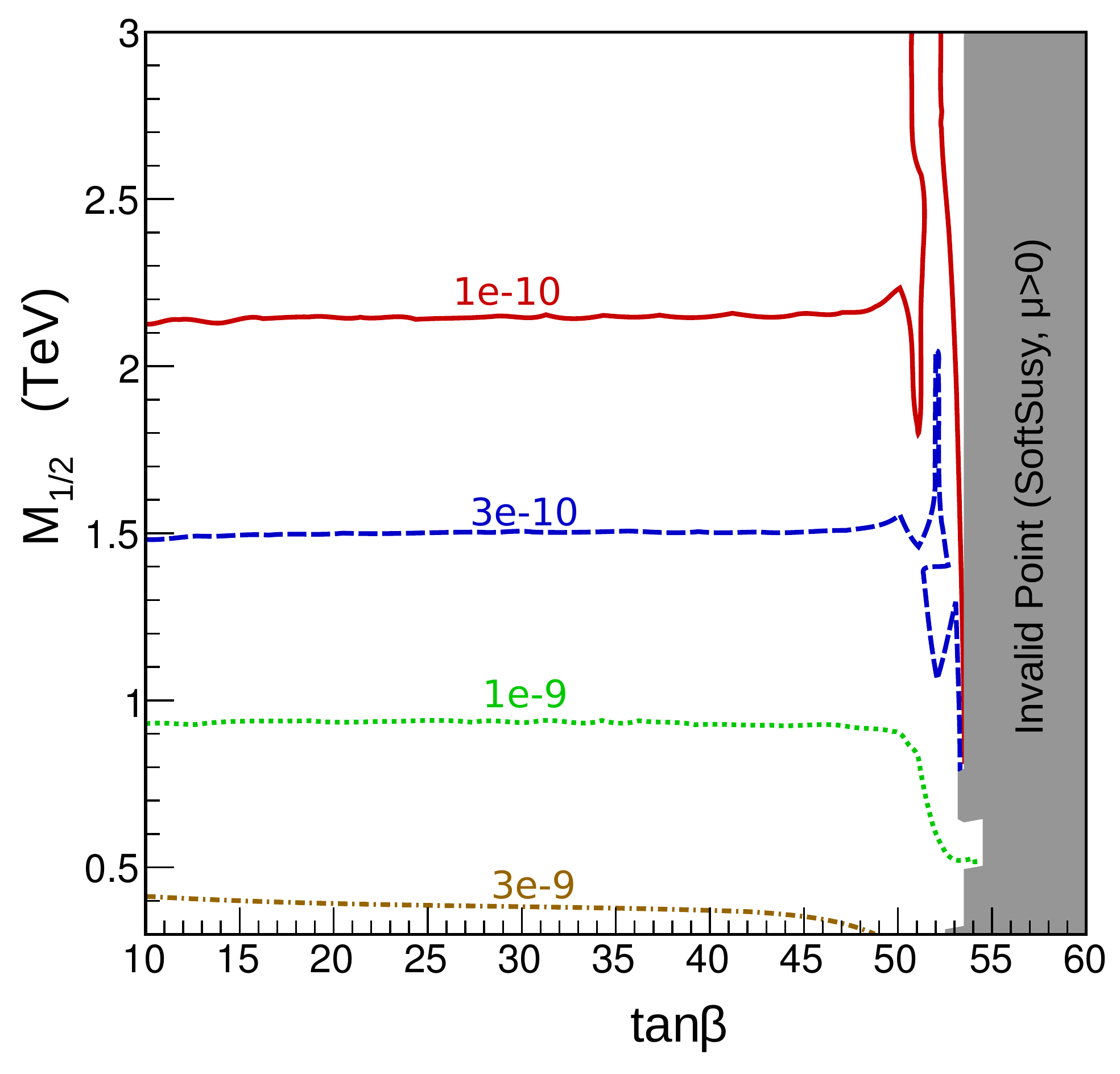}
\label{fig:mupospbar}
}
\caption{The differential flux of anti-protons in units of
  $\gev^{-1}~\cm^{-2}~\s^{-1}~\sr^{-1}$ at an energy of 19.6 GeV.}
   \label{fig:pbar}
\end{figure}

\Figref{pbar} shows the flux of anti-protons at an energy of 19.6 GeV.
We use the default propagation parameters for charged cosmic rays in
DarkSUSY, as well as the default DarkSUSY~\cite{ds} dark matter halo
density profile.  We choose the particular energy of 19.6 GeV for two
reasons:

\noindent (1) It was shown in Ref.~\cite{Baer:2005ky} (see Fig.~10,
left and Fig.~11, left) that the best signal to background ratio for
anti-proton searches in the focus point region ranges between 10 and
100 GeV in kinetic energy, with an optimal value of about 20 GeV when
factoring in the need to observe a large enough number of signal
events;

\noindent (2) 19.6 GeV corresponds to the central value of the
relevant energy bin reported by the PAMELA
Collaboration~\cite{Adriani:2010rc}. At that energy, PAMELA quotes a
flux of $7.2\times 10^{-8}~\gev^{-1}~\cm^{-2}~\s^{-1}~\sr^{-1}$.

The contours in \figref{pbar} indicate that the signal-to-noise ratio
expected in the ``sweet spot'' for the anti-proton kinetic energy
ranges between 2\% for very light neutralinos to less than 0.1\% for
more massive neutralinos.  We find almost no variation between
negative and positive values of $\mu$. Given the absence of any
striking spectral feature in the predicted spectrum of anti-protons in
the focus point region~\cite{Baer:2005ky}, and the fact that
variations in the cosmic ray anti-proton diffusion and energy loss
parameters can induce deformation to the background spectrum much
larger than the percent level, we conclude that the predicted flux of
anti-protons is generically too small to provide a conclusive dark
matter detection avenue for neutralinos in the focus point region.

\begin{figure}[tb]
\subfigure[ \ $\mu < 0$]{
\includegraphics[width=0.48\textwidth]{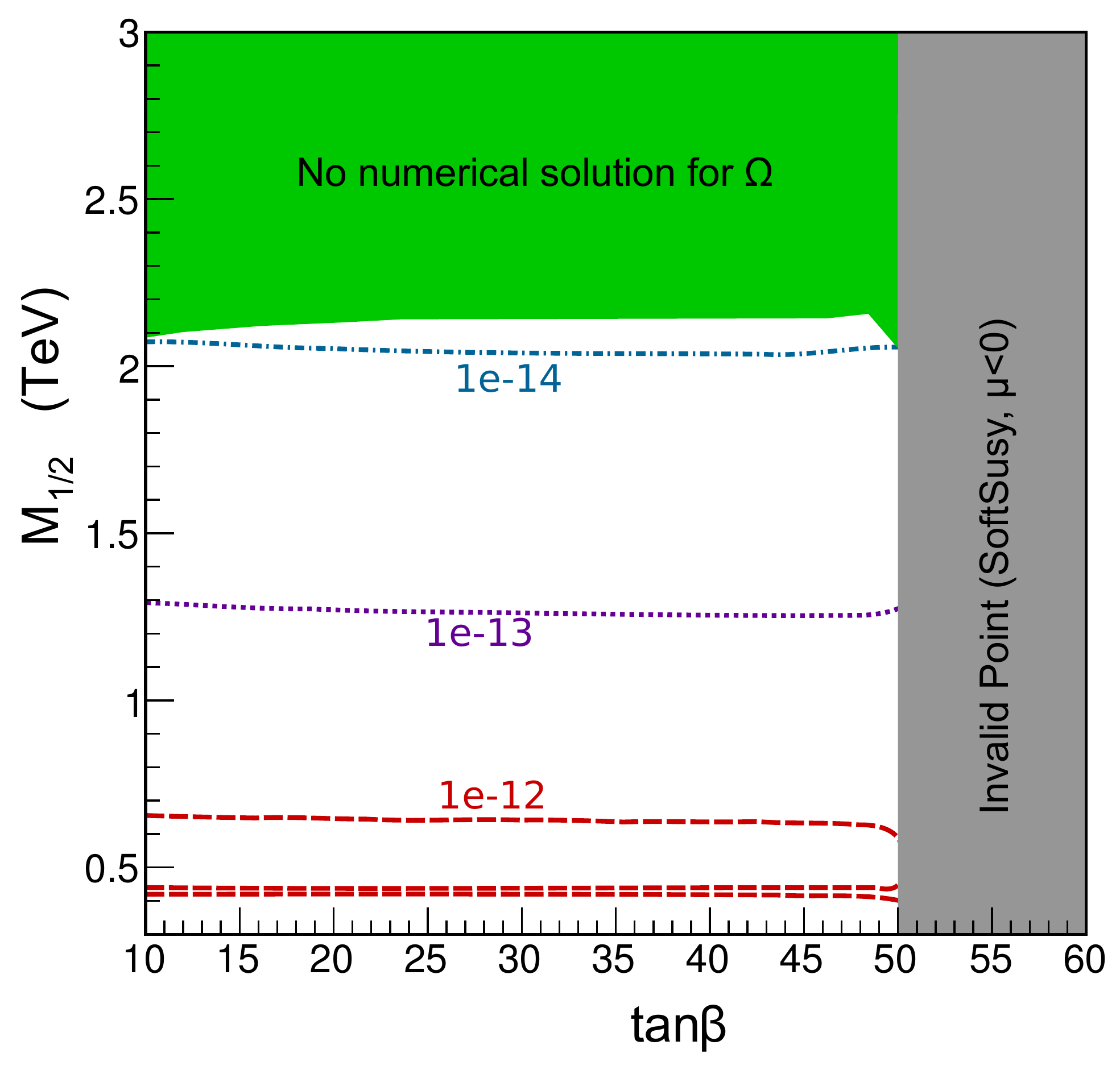}
\label{fig:munegdbar}
}
\subfigure[ \ $\mu > 0$]{
\includegraphics[width=0.48\textwidth]{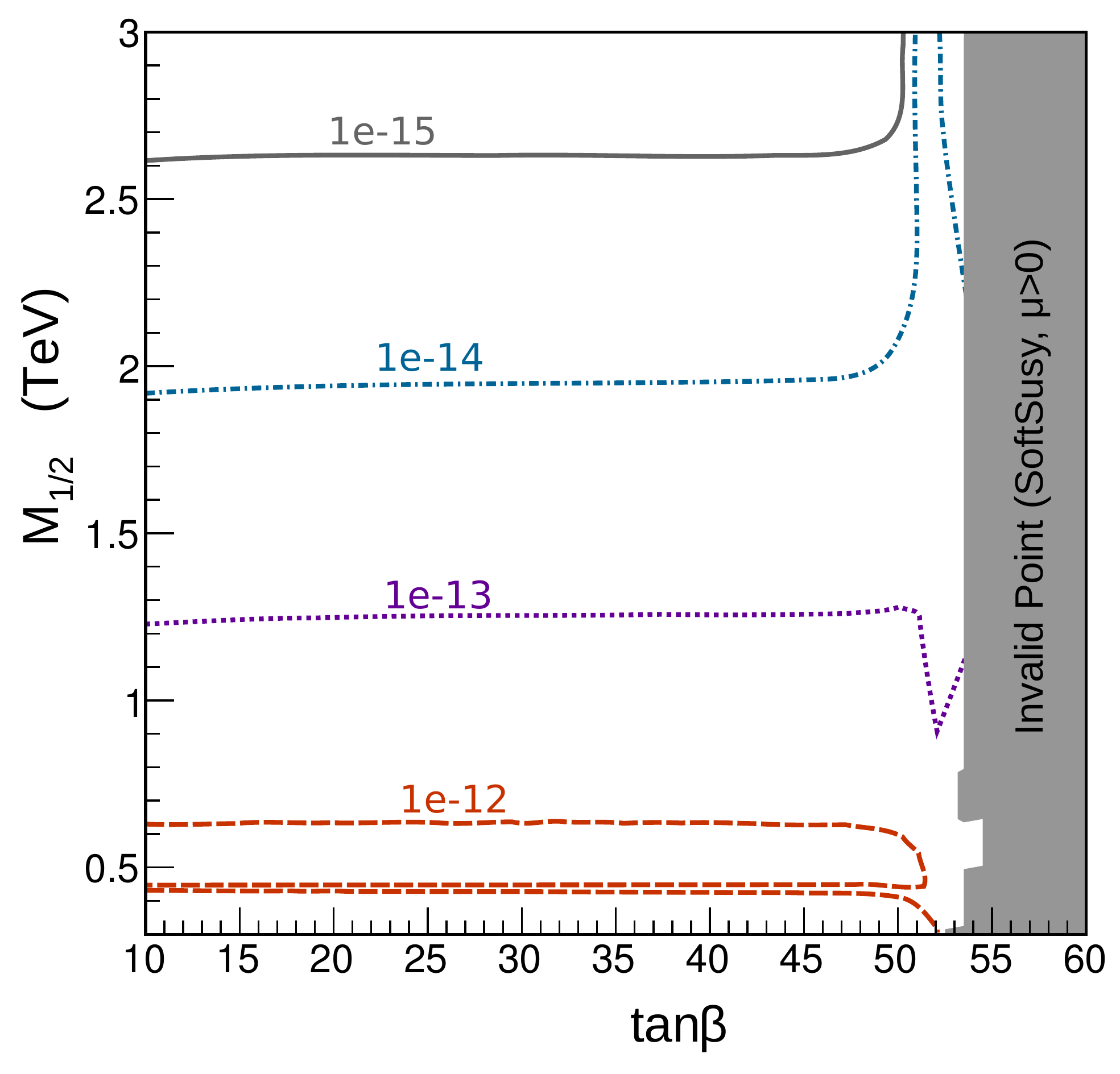}
\label{fig:muposdbar}
}
\caption{The flux of anti-deuterons at an energy of 1 GeV, in units of
  $\gev^{-1}~\cm^{-2}~\s^{-1}~\sr^{-1}$.}
\label{fig:dbar}
\end{figure}

\Figref{dbar} shows the anti-deuteron flux at 1 GeV. Although the GAPS
experiment will primarily target lower energies (likely between 0.1
and 0.3 GeV), the AMS-02 limits are likely to be best in the 1 GeV
range. In addition, the predicted flux at 0.1--0.3 GeV is typically
comparable (within less than a factor 2) to that at 1 GeV for
neutralinos in the focus point region.

The GAPS experimental sensitivity target is at present estimated to be
at the level of just under $10^{-11}$ GeV$^{-1}$ cm$^{-2}$ s$^{-1}$
sr$^{-1}$, while AMS-02 should be able to reach a sensitivity of about
$10^{-10}$ GeV$^{-1}$ cm$^{-2}$ s$^{-1}$ sr$^{-1}$, or approximately
an order of magnitude less constraining than GAPS. \Figref{dbar}
therefore illustrates that across the relevant focus point region
parameter space the expected anti-deuteron signal is between 1 and 4
orders of magnitude smaller than the best foreseeable experimental
sensitivity, making this indirect detection channel inconclusive to
search for a dark matter signal.

\section{Discussion and Conclusions}
\label{sec:concl}

In the context of supersymmetric extensions of the standard model, the
discovery of a relatively heavy Higgs boson at the LHC, coupled with
null results from superpartner searches, provides strong motivation
for considering models with multi-TeV squarks and sleptons.  We
consider cosmologically-motivated focus point model realizations of
this scenario in which dark matter is entirely composed of thermal
relics that are mixed Bino-Higgsino or Higgsino-like neutralinos.

Our main findings are the following:
\begin{itemize}
\item These models remain viable.  Claims to the contrary are
  apparently the result of (a) requiring supersymmetry to resolve the
  $(g-2)_{\mu}$ discrepancy (a requirement that is tantamount to
  considering the standard model to be excluded by this discrepancy),
  (b) considering only $\mu>0$ (presumably for historical reasons
  linked to (a)), (c) using large values of $f_s$ that are now highly
  disfavored, (d) imposing some highly subjective naturalness
  criterion, or (e) a combination of these.
\item The leading 3-loop ${\cal O} ( \alpha_t \alpha_s^2)$
  contributions to the Higgs mass are positive, lowering the preferred
  values of scalar masses (possibly to values within reach of the LHC)
  and improving the fine-tuning of these scenarios.
\item Some focus point parameter space is excluded by bounds from
  direct searches for dark matter, but some remains, including much of
  the parameter space with $\mu < 0$.  In the allowed regions, the
  predicted spin-independent cross sections are just beyond current
  bounds from XENON, and spin-dependent scattering is also close to
  the experimental sensitivity expected in the near future.
\item For indirect detection, searches for neutrinos from the core of
  the Sun at IceCube/DeepCore exclude focus point neutralinos lighter
  than about 170 GeV.  The anticipated detector performance would have
  placed constraints on neutralinos as heavy as 600 GeV, covering most
  of the focus point parameter space.  There are therefore bright
  prospects for dark matter discovery through neutrinos at
  IceCube/DeepCore. Similar sensitivity is being reached by other
  experiments, such as ANTARES.  These results are insensitive to halo
  model choices, and also do not depend on, e.g., the strange quark
  content of the proton, and so yield promising probes that are highly
  complementary to direct detection.
\item The predicted neutrino flux from the center of the Earth is many
  orders of magnitude below detectability.
\item We have also considered gamma rays from the galactic center and
  dwarf galaxies producing either line or continuum signals.  Signals
  in gamma rays are not as promising as in neutrinos, at least for the
  conservative choices of the relevant dark matter density profiles we
  employed here, but may nevertheless still be seen in future
  experiments such as CTA.
\item For indirect detection of anti-protons, the signal-to-background
  ratio, even at optimal energies, is at the percent level and too
  small to provide a convincing avenue for dark matter detection.
\item Anti-deuteron rates are one to four orders of magnitude below
  the foreseen experimental sensitivity of future dedicated
  experiments, such as GAPS.
\end{itemize}

To summarize, LHC results so far motivate focus point supersymmetry,
which has exciting implications for dark matter searches. Among the
most promising are direct searches for spin-independent scattering and
indirect searches with neutrino telescopes, but other approaches
discussed here may also yield signals.  Uncertainties in the Higgs
mass calculation also leave open the possibility that squarks and
gluinos may be within reach of the LHC, even without large left-right
stop mixing.  If focus point supersymmetry is realized in nature and
focus point neutralinos make up all of the dark matter in the
Universe, a signal in one or more of the complementary probes
(colliders, direct detection, and indirect detection) will appear in
the coming few years.

\section*{Acknowledgments}

PD and SP are partly supported by the US Department of Energy under
Contract No.~DE--FG02--04ER41268.  JLF is supported in part by NSF
Grant No.~PHY--0970173 and a Simons Foundation Fellowship.  DS is
supported in part by the U.S. Department of Energy under contract
No.~DE--FG02--92ER40701 and by the Gordon and Betty Moore Foundation
through Grant No.~776 to the Caltech Moore Center for Theoretical
Cosmology and Physics.  PK is supported by the DFG through SFB/TR-9
and by the Helmholtz Alliance ``Physics at the Terascale.''

\bibliography{bibfpdm2}{}

\end{document}